\documentclass[preprint,nofootinbib,tightenlines,amsfonts,amsmath,amssymb]{revtex4}
\usepackage{bm}% bold math
\usepackage{color}
\usepackage{graphicx}% Include figure files

\begin{document}
\baselineskip=15pt \parskip=5pt

\vspace*{3em}

\preprint{NCTS-PH/1707}

\title{Two-Higgs-Doublet-Portal Dark-Matter Models\\in Light of Direct Search and LHC Data}

\author{Chia-Feng Chang$^{1}$}
\author{Xiao-Gang He$^{1,2,3}$}
\author{Jusak Tandean$^{1,3}$}

\affiliation{$^1$Department of Physics and Center for Theoretical Sciences, National Taiwan
University,\\ No.\,\,1, Sec.\,\,4, Roosevelt Rd., Taipei 106, Taiwan \smallskip \\
$^2$INPAC, Department of Physics and Astronomy, Shanghai Jiao Tong University,
800 Dongchuan Rd., Minhang, Shanghai 200240, China \smallskip \\
$^3$Physics Division, National Center for Theoretical Sciences,
No.\,\,101, Sec.\,\,2, Kuang Fu Rd., Hsinchu 300, Taiwan \bigskip}

%\date{$\vphantom{\bigg|_{\bigg|}^|}$}

\begin{abstract}
We explore simple Higgs-portal models of dark matter (DM) with spin 1/2, 3/2, and 1,
respectively, applying to them constraints from the LUX and PandaX-II direct detection
experiments and from LHC measurements on the 125-GeV Higgs boson.
With only one Higgs doublet, we find that the spin-1/2 DM having a purely scalar effective
coupling to the doublet is viable only in a narrow range of mass near the Higgs pole,
whereas the vector DM is still allowed if its mass is also close to the Higgs pole or
exceeds 1.4 TeV, both in line with earlier analyses.
Moreover, the spin-3/2 DM is in a roughly similar situation to the spin-1/2 DM, but has
surviving parameter space which is even more restricted.
We also consider the two-Higgs-doublet extension of each of the preceding models, assuming
that the expanded Yukawa sector is that of the two-Higgs-doublet model of type II.
We show that in these two-Higgs-doublet-portal models significant portions of the DM mass
regions excluded in the simplest scenarios by direct search bounds can be reclaimed due
to suppression of the effective DM interactions with nucleons at some ratios of the
$CP$-even Higgs bosons' couplings to the up and down quarks.
The regained parameter space contains areas which can yield a DM-nucleon scattering
cross-section that is far less than its current experimental limit or even goes below
the neutrino-background floor.
\end{abstract}

\maketitle

\section{Introduction\label{intro}}

The latest direct searches for weakly interacting massive particle (WIMP) dark matter (DM)
by the LUX and PandaX-II Collaborations~\cite{lux,pandax} have turned up null results,
leading to the strictest upper-limits to date on the cross section of spin-independent
elastic WIMP-nucleon scattering in the mass region from about 4 GeV to 100 TeV.
For lower WIMP masses down to 0.5 GeV, the existing corresponding limits were set a little
earlier in the CRESST~\cite{Angloher:2015ewa} and CDMSlite~\cite{cdmslite} experiments.
Ongoing and planned efforts to detect the DM directly~\cite{Cushman:2013zza}
will likely improve upon these findings if it still eludes discovery.

The above recent measurements translate into substantial restrictions on WIMP DM models,
especially minimal Higgs-portal ones, which are also subject to constraints
from continuing quests at the LHC~\cite{Aad:2015pla,atlas+cms} for decays of the 125-GeV Higgs
boson into final states that would signal new physics beyond the standard model (SM).
The combination of restraints from direct detection and LHC data has ruled out in particular
the DM mass region below 0.5 TeV or so in the simplest Higgs-portal real-scalar-DM model,
with the exception of a narrow range around the Higgs
pole~\cite{He:2016mls,Wu:2016mbe,Escudero:2016gzx}.
If the DM is instead a\,\,spin-1/2 fermion with a purely scalar effective coupling
to the Higgs, only this small region near the Higgs pole remains
viable~\cite{Escudero:2016gzx,Beniwal:2015sdl}.

Nevertheless, as previously demonstrated in the Higgs-portal scalar-DM
case~\cite{He:2016mls,He:2008qm,Greljo:2013wja}, by adding another Higgs doublet to
the simplest scenario it is possible to decrease the effective interactions of the DM with
nucleons sufficiently and thereby to regain at least some of the parameter space disallowed
by the direct searches.
This motivates us to explore similar ideas in other simple models.
Specifically, in this paper we first revisit the minimal Higgs-portal scenarios in which
the DM is a fermion of spin 1/2 or 3/2 or a spin-1 boson, to see how the aforesaid
restraints impact them.
Subsequently, we consider a\,\,somewhat expanded version of each of the models by
incorporating another Higgs doublet and arranging the new Yukawa sector to be that of
the so-called two-Higgs-doublet model (THDM) of type II.
We will show that these two-Higgs-doublet-portal DM models, like their scalar-DM
counterpart, potentially have ample parameter space that can avoid all the latest
direct-detection limits and may even evade future ones.
It is worth mentioning here that there have been a number of studies in the past on various
Higgs-portal scenarios in which a THDM was supplemented with a SM-gauge-singlet DM candidate
having spin 0~\cite{He:2008qm,Greljo:2013wja,He:2011gc,2hdm+d,Gao:2011ka,Cai:2013zga},
1/2~\cite{Gao:2011ka,Cai:2013zga,Banik:2013zla}, or 1~\cite{Cai:2013zga}.

The organization of the rest of the paper is as follows.
In Sec.\,\,\ref{sec:sm+fdm} we take another look at the available constraints from DM direct
searches and LHC Higgs measurements on the three different minimal Higgs-portal models
having WIMP DM candidates with spin 1/2, 3/2, and\,\,1, respectively.
In Sec.\,\,\ref{sec:2hdm+fdm}, we deal with the two-Higgs-doublet extensions of these models
where the enlarged Yukawa sector is that of the type-II THDM.
We will address how the extended models can escape some of the restrictions in the presence of
sizable breaking of isospin symmetry in the DM effective interactions with
nucleons.\footnote{Isospin violation in DM interactions can occur not only in a THDM plus
SM-singlet DM~\cite{He:2016mls,He:2008qm,Greljo:2013wja,He:2011gc,2hdm+d,Gao:2011ka,Cai:2013zga,
Banik:2013zla}, but also in certain other models, such as those in which the DM couples to
a $Z'$ boson~\cite{Frandsen:2011cg}.
More general aspects of isospin-violating DM have been discussed
in~\cite{Kurylov:2003ra,Chang:2010yk}.}
We give our conclusions in Sec.\,\,\ref{conclusion}.
We collect additional formulas and extra details in a couple of appendices.

\section{Minimal Higgs-portal fermionic and vector DM models\label{sec:sm+fdm}}

\subsection{Spin-1/2 dark matter\label{half}}

In the most economical fermionic scenario~\cite{Kim:2006af}, the SM is slightly enlarged by
the inclusion of a spin-1/2 Dirac field $\psi$ which is a singlet under the SM gauge
group and serves as the WIMP DM candidate.
In this model, hereafter referred to as SM+$\psi$, the DM is stable due to an exactly
preserved $Z_2$ symmetry under which \,$\psi\to-\psi$,\, while the SM fields are unchanged.
Since $\psi$ cannot couple directly to SM members in a renormalizable way,
without explicitly introducing other new ingredients one can explore $\psi$ interactions
with the SM sector that are induced by effective nonrenormalizable operators.
The simplest ones with a Higgs doublet $\sf H$ are the dimension-five combinations
\,$\overline{\psi}(1\pm\gamma_5)\psi\,{\sf H}^\dagger{\sf H}$,\, which are invariant under
the SM gauge group.
Assuming that the Lagrangian ${\cal L}_\psi$ for $\psi$ conserves $CP$ symmetry,
making the pseudoscalar coupling absent, one can then write~\cite{Kim:2006af}
\begin{eqnarray} \label{Lpsi0}
{\cal L}_\psi^{} \,=\, \overline{\psi}\,i\slash{\!\!\!\partial}\psi \,-\,
\mu_\psi^{}\overline{\psi}_{\,\!}\psi \,-\,
\frac{\overline{\psi}\psi\,{\sf H}^\dagger{\sf H}}{\Lambda_\psi^{}} \,,
\end{eqnarray}
where $\mu_\psi$ and $\Lambda_\psi$ are real constants of dimension mass and $\Lambda_\psi$
absorbs the parameters of the underlying heavy physics.\footnote{Further phenomenology of
the DM described in Eq.\,(\ref{Lpsi0}) has been explored before in~\cite{Beniwal:2015sdl,
Low:2011kp,Kanemura:2010sh,Djouadi:2011aa,Kamenik:2011vy,Fedderke:2014wda,LopezHonorez:2012kv}.
Different possibilities for its ultraviolet (UV) completion have also been proposed
in~\cite{LopezHonorez:2012kv,Kim:2008pp,Baek:2014jga}.}
After electroweak symmetry breaking, \,${\sf H}^\dagger{\sf H}=\tfrac{1}{2}(h+v)^2$,\, and so
\begin{eqnarray} \label{Lpsi}
{\cal L}_\psi^{} \,\supset\, - m_\psi^{}\overline{\psi}_{\,\!}\psi
- \lambda_{\psi h}^{}\,\overline{\psi}\psi\,\bigg(h+\frac{h^2}{2v}\bigg) \,, ~~~~ ~~~
m_\psi^{} \,=\, \mu_\psi^{}+\frac{\lambda_{\psi h}^{}v}{2} \,, ~~~~ ~~~
\lambda_{\psi h}^{} \,=\, \frac{v}{\Lambda_\psi^{}} \,,
\end{eqnarray}
where $h$ is the physical Higgs and \,$v\simeq246$\,GeV\, the vacuum expectation
value (VEV) of $\sf H$.
The DM mass $m_\psi$ and the DM-Higgs coupling $\lambda_{\psi h}$
are the only free parameters in ${\cal L}_\psi$.

The $\lambda_{\psi h}$ terms in Eq.\,(\ref{Lpsi}) are responsible for the DM relic density.
It results from $\bar\psi\psi$ annihilation into SM particles, which happens mainly
via the Higgs-exchange process \,$\bar\psi\psi\to h^*\to X_{\textsc{sm}}$.\,
If the $\bar\psi\psi$ pair has a center-of-mass (c.m.) energy \,$\sqrt s>2m_h^{}$,\,
the channel \,$\bar\psi\psi\to hh$\, needs to be taken into account.
Thus, we can express the cross section $\sigma_{\rm ann}$ of DM annihilation as
\begin{eqnarray}
\sigma_{\rm ann}^{} &=& \sigma\big(\bar\psi\psi\to h^*\to X_{\textsc{sm}}\big) \,+\,
\sigma\big(\bar\psi\psi\to hh\big) \,,
\vphantom{|_{\int_|^|}^{}} \nonumber \\ \label{DD2h2sm}
\sigma\big(\bar\psi\psi\to h^*\to X_{\textsc{sm}}\big) &=&
\frac{\beta_{\psi\,}^{}\lambda_{\psi h\,}^2\sqrt s~\raisebox{3pt}
{\footnotesize$\displaystyle\sum_i$}\,\Gamma\big(\tilde h\to X_{i,\textsc{sm}}\big)}
{2\big[\big(m_h^2-s\big)\raisebox{1pt}{$^2$}+\Gamma_h^2m_h^2\big]} \,, ~~~~ ~~~
\beta_{\textsc x}^{} \,=\, \sqrt{1-\frac{4m_{\textsc x}^2}{s}} ~, ~~~ ~~~~
\end{eqnarray}
where the formula for $\sigma\big(\bar\psi\psi\to hh\big)$ is relegated to
Appendix\,\,\ref{formulas}, the sum in the second line is over SM final-states
\,$X_{i,\textsc{sm}}\neq hh$,\, and $\tilde h$ refers to a virtual Higgs having a mass
\,$m_{\tilde h}=\sqrt s$.\,
Subsequently, we can extract $\lambda_{\psi h}$ from the observed DM abundance, as outlined
in Appendix\,\,\ref{formulas}, and then test the result with various constraints.

One of the important restrictions on $\lambda_{\psi h}$ applies
in the region \,$m_\psi^{}<m_h^{}/2$,\, where the invisible decay channel
\,$h\to\bar\psi\psi$\, is open and contributes to the Higgs' total width
\,$\Gamma_h^{}=\Gamma_h^{\textsc{sm}}+\Gamma(h\to\bar\psi\psi)$.\,
From Eq.\,(\ref{Lpsi}), we obtain the partial rate
\begin{eqnarray} \label{Gh2DD}
\Gamma\big(h\to\bar\psi\psi\big) \,=\, \frac{\lambda_{\psi h\,}^2m_h^{}}{8\pi}
\Bigg(1-\frac{4 m_\psi^2}{m_h^2}\Bigg)^{\!3/2} \,.
\end{eqnarray}
The experiments at the LHC offer information pertaining to this process.
According to the latest combined analysis by the ATLAS and CMS Collaborations on their
Higgs measurements~\cite{atlas+cms}, the branching fraction of $h$ decay into channels
beyond the SM is \,${\cal B}_{\textsc{bsm}}^{\rm exp}=0.00^{+0.16}$,\, which can be
interpreted as placing a cap on the invisible decay of $h$, explicitly
\,${\cal B}(h\to\rm invisible)_{\rm exp}<0.16$.\,
Consequently, we can demand
\begin{eqnarray} \label{lhclimit}
{\cal B}\big(h\to\bar\psi\psi\big) \,=\,
\frac{\Gamma\big(h\to\bar\psi\psi\big)}{\Gamma_h^{}} \,<\, 0.16 \,.
\end{eqnarray}
In numerical work, we fix \,$m_h^{}=125.1$\,GeV,\, based on the current data~\cite{lhc:mh},
and correspondingly the SM Higgs width \,$\Gamma_h^{\textsc{sm}}=4.08$\,MeV \cite{lhctwiki}.

Another major constraint on $\lambda_{\psi h}$ is supplied by direct detection experiments,
which look for nuclear recoil effects caused by the DM colliding with a nucleon, $N$,
nonrelativistically at momentum transfers small relative to the nucleon mass, $m_N$.
In the SM+$\psi$, this is an elastic transition, \,$\psi N\to\psi N$,\, which is mediated
by the Higgs in the $t$ channel.
Its cross section is
\begin{eqnarray} \label{csel}
\sigma_{\rm el}^N \,=\, \frac{\lambda_{\psi h\,}^2g_{NNh\,}^2m_\psi^2m_N^2}
{\pi\,\big(m_\psi^{}+m_N^{}\big)\raisebox{1pt}{$^2$}m_h^4} \,.
\end{eqnarray}
where $g_{NNh}^{}$ is the effective Higgs-nucleon coupling.
Numerically, we adopt \,$g_{NNh}^{}=0.0011$\, following Ref.\,\cite{He:2016mls}.
The strongest restraints on $\sigma_{\rm el}^N$ to date for
\,$m_\psi\mbox{\footnotesize$\,\gtrsim\,$}5$\,GeV\, are provided by
LUX~\cite{lux} and PandaX-II~\cite{pandax}.

Given that \,$\Lambda_\psi^{-1}=\lambda_{\psi h}/v$\, in Eq.\,(\ref{Lpsi}) is
the coefficient of a dimension-5 effective operator, the size of $\lambda_{\psi h}$ is also
capped by the extent of validity of the effective field theory (EFT) description for
the $\psi$-{\sf H} interactions.
To derive a rough estimate on the minimum of $\Lambda_\psi$, we entertain the possibility
that this operator is induced by a tree-level diagram mediated by a heavy scalar $X$ with
mass $m_X^{}$ and couplings to $\psi$ and $h$ described by
\,${\cal L}_X^{}\supset-g_\psi^{}\bar\psi\psi X-g_h^{}h^2X$\, in the UV-complete theory.
In addition, we suppose that \,$g_\psi^{}\sim m_\psi^{}/v_X^{}$\, and
\,$g_h^{}\sim\lambda_{hX}^{}v_X^{}$,\, where $v_X^{}$ is the VEV of $X$ and $\lambda_{hX}$ is
dimensionless, inspired by the fermionic and scalar couplings in the\,\,SM,
ignoring potential modifications due to $h$-$X$ mixing.
The EFT will then remain a good approximation and perturbative if
\,$1/|\Lambda_\psi|\sim 2|\lambda_{hX}|m_\psi/m_X^2<|\lambda_{hX}|/(2m_\psi)<2\pi/m_\psi$,\,
as the $s$-channel $\bar\psi\psi$ energy $\sqrt s$ satisfies \,$m_X^2>s>4m_\psi^2$\, and
\,$|\lambda_{hX}|<4\pi$\, for perturbativity.\footnote{The same bound on $\Lambda_\psi$ was
obtained in \cite{Beniwal:2015sdl,Busoni:2014sya} using similar arguments.}
We then have \,$|\lambda_{\psi h}|<2\pi v/m_\psi^{}$.\,
As this follows from the most relaxed requisite on $\lambda_{hX}$, it is
likely that the EFT description breaks down at a significantly smaller $\lambda_{\psi h}$.
Therefore, alternatively it is reasonable to set \,$|\lambda_{hX}|<2$,\, leading to
\,$|\lambda_{\psi h}|<v/m_\psi^{}$.\,
In the \,$m_\psi<m_h^{}/2$\, region, this restriction turns out to be far weaker than
that from Eq.\,(\ref{lhclimit}) for the Higgs invisible decay, as will be seen shortly.

\begin{figure}[b]
\includegraphics[width=73mm]{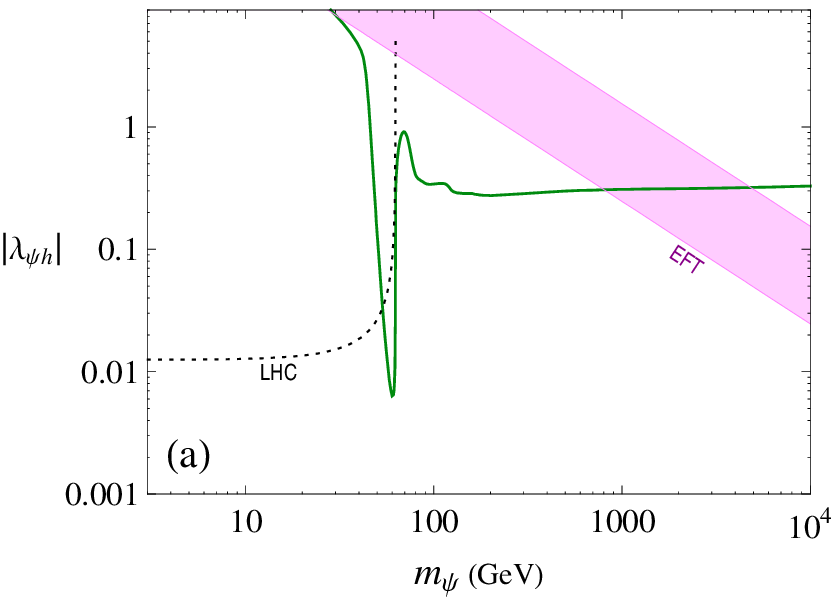} ~ ~
\includegraphics[width=75.3mm]{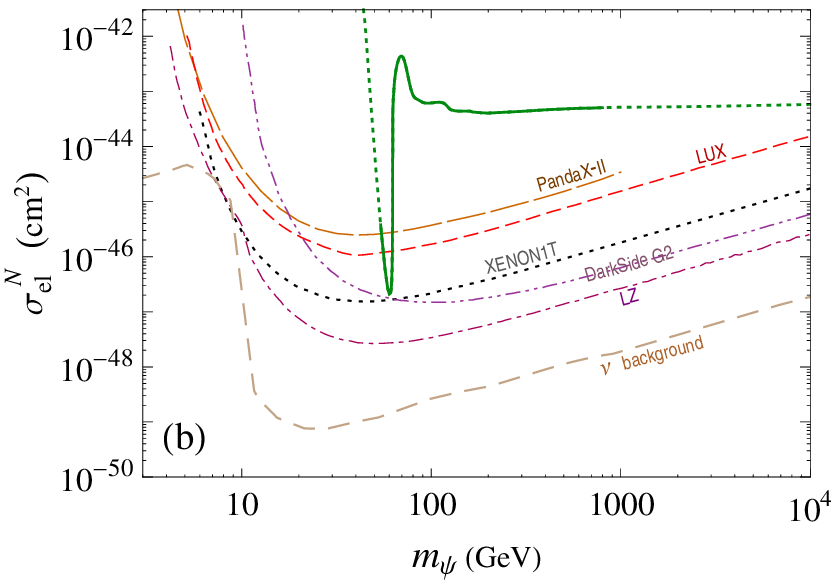}\vspace{-7pt}
\caption{(a) The magnitude of DM-Higgs coupling $\lambda_{\psi h}$ versus DM mass $m_\psi$
in the SM+$\psi$ satisfying the relic density requirement (green curve), compared to
the upper limits inferred from LHC data on Higgs invisible decay (black dotted curve) and
from the validity extent of the EFT approximation (lower side of magenta band, as discussed
in the text).
(b)\,\,The corresponding cross-section~$\sigma_{\rm el}^{N}$ of $\psi$-nucleon elastic
scattering (green curve), compared to the measured upper-limits at 90\% confidence level
from LUX~\cite{lux} (red dashed curve) and PandaX-II~\cite{pandax} (orange long-dashed curve).
Also shown are the sensitivity projections~\cite{Cushman:2013zza} of XENON1T~\cite{xenon1t}
(black dotted curve), DarkSide\,\,G2~\cite{dsg2} (purple dash-dot-dotted curve), and
LZ~\cite{lz} (maroon dash-dotted curve), and the WIMP discovery lower-limit due to coherent
neutrino scattering backgrounds~\cite{nubg} (brown dashed curve).
The dotted portions of the green curve are excluded by the LHC and EFT restrictions in (a).}
\label{sm+fdm_half-plots} \vspace{-1ex}
\end{figure}

To illustrate how the model confronts these different requirements, we display in
Fig.\,\ref{sm+fdm_half-plots}(a) the values of $|\lambda_{\psi h}|$ derived from the observed
DM relic abundance (green solid curve) and compare them to the upper limits on
$|\lambda_{\psi h}|$ inferred from Eq.\,(\ref{lhclimit}) based on LHC data~\cite{atlas+cms}
(black dotted curve) and from the validity extent of the EFT approach.
For the latter, based on the discussion in the preceding paragraph we draw
the magenta band corresponding to the region \,$|\lambda_{\psi h}|\in[1,2\pi]v/m_\psi^{}$\,
in which the EFT description may be expected to have broken down.
Thus, we can regard the lower boundary of this band as the upper limit for the reliability
of the EFT approximation.
We determine the range allowed by these constraints to be
\,$54{\rm\;GeV}\mbox{\footnotesize\,$\lesssim$\,}m_\psi^{}\mbox{\footnotesize\,$\lesssim$\,}
0.8\;$TeV,\,
which translates into the solid portion of the green curve in Fig.\,\ref{sm+fdm_half-plots}(b)
for the $\psi$-$N$ cross-section, $\sigma_{\rm el}^N$.
This green solid curve turns out to be forbidden by the LUX bound,
except in a slender zone near the Higgs pole, more precisely
\,55.8\,\,GeV\,$\mbox{\footnotesize$\lesssim$}\,m_\psi\,
\mbox{\footnotesize$\lesssim$}\;$62.3\,\,GeV.\,
Similar results were found in Refs.\,\cite{Beniwal:2015sdl,Escudero:2016gzx}.
The ongoing PandaX-II as well as the planned XENON1T, DarkSide G2, and LZ
experiments~\cite{Cushman:2013zza} will likely be able to probe the surviving parameter
space exhaustively.

It is worth remarking that the $\lambda_{\psi h}$ values in Fig.\,\ref{sm+fdm_half-plots}(a)
are much bigger than most of their counterparts in the simplest scalar-DM
model~\cite{He:2016mls,Wu:2016mbe,Escudero:2016gzx}.
This enlargement is compensation for the suppression of the DM annihilation rate,
$\sigma_{\rm ann}v_{\rm rel}$, by two powers of the c.m. relative speed $v_{\rm rel}$ of
$\bar\psi$ and $\psi$ in the nonrelativistic limit, as can be easily
checked.\footnote{If $CP$ invariance is not imposed on ${\cal L}_\psi$ in Eq.\,(\ref{Lpsi0}),
it can accommodate the combination \,$\overline{\psi}\gamma_5^{}\psi\,{\sf H}^\dagger{\sf H}$.
After electroweak symmetry breaking, this operator generally gives rise to both scalar and
pseudoscalar contributions to DM annihilation, \,$h\to\bar\psi\psi$, and DM-nucleon
scattering~\cite{Beniwal:2015sdl,Fedderke:2014wda}.
The pseudoscalar one can alleviate the $v_{\rm rel}^2$ suppression in
\,$\sigma_{\rm ann}v_{\rm rel}$,\, but yields a tiny effect on the DM-nucleon cross-section.}
As a consequence, the enhanced prediction for $\sigma_{\rm el}^N$ is in conflict with
the LUX bound over a wider mass region than in the scalar-DM case.

\subsection{Spin-\boldmath3/2 dark matter\label{rs}}

The WIMP DM in this scenario is described by a Rarita-Schwinger field~\cite{rs} which is
denoted here by a Dirac four-spinor $\Psi_\nu$ with a vector index $\nu$ and satisfies
the relation
\,$\gamma^\nu\Psi_\nu=0$.\footnote{The basic properties of this kind of spin-3/2 fermion,
especially in the DM context, have been elaborated in~\cite{Kamenik:2011vy,Yu:2011by}.\medskip}
In the minimal model, called SM+$\Psi$, the DM is a SM-gauge singlet, its stability is
maintained by an unbroken $Z_2$ symmetry under which \,$\Psi_\nu\to-\Psi_\nu$,\, the SM fields
being unaffected, and the Higgs-portal interactions arise from dimension-5
operators~\cite{Kamenik:2011vy}, like in the SM+$\psi$.
The DM Lagrangian, assumed again to be $CP$ invariant, is then
\begin{eqnarray} \label{LP0}
{\cal L}_{\Psi}^{} \,=\,
-\overline{\Psi_\nu}\big(i\slash{\!\!\!\partial}-\mu_{\Psi}^{}\big)\Psi^\nu
+ \frac{\overline{\Psi_\nu}_{\,}\Psi^\nu\,{\sf H}^\dagger{\sf H}}{\Lambda_{\Psi}^{}}
\,\supset\,
m_{\Psi}^{}\,\overline{\Psi_\nu}_{\,}\Psi^\nu
+ \lambda_{\Psi h\,}^{}\overline{\Psi_\nu}_{\,}\Psi^\nu\,\bigg(h+\frac{h^2}{2v}\bigg) \,,
\end{eqnarray}
where $\mu_{\Psi}$ and $\Lambda_{\Psi}$ are real constants, $\Lambda_{\Psi}$ encodes
the underlying heavy physics, and the DM mass and coupling
\begin{eqnarray}
m_{\Psi}^{} \,=\, \mu_{\Psi}^{}
+ \frac{\lambda_{\Psi h\,}^{}v}{2} \,, ~~~~ ~~~
\lambda_{\Psi h}^{} \,=\, \frac{v}{\Lambda_{\Psi}^{}} \,,
\end{eqnarray}
respectively, are the only free parameters in ${\cal L}_\Psi$.

Accordingly, we can derive the main quantities relevant to the DM phenomenology in analogy
to the spin-1/2 case.
Thus, the DM-annihilation cross-section $\sigma_{\rm ann}$ is given
by\footnote{The $\bar{\Psi}\Psi$ annihilation rate, $\sigma_{\rm ann}^{}v_{\rm rel}^{}$, like
its spin-1/2 counterpart, suffers from $v_{\rm rel}^2$ suppression in the nonrelativistic limit.
If ${\cal L}_{\Psi}$ is not $CP$-invariant, it can include the pseudoscalar
combination \,$\overline{\Psi_\nu}\gamma_5^{}\Psi^\nu\,{\sf H}^\dagger{\sf H}$.\,
In its presence, there is generally an admixture of scalar and pseudoscalar contributions
to $\Psi$-$h$ interactions which can ameliorate the $v_{\rm rel}^2$ suppression
in $\sigma_{\rm ann}^{}v_{\rm rel}^{}$.}
\begin{eqnarray}
\sigma_{\rm ann}^{} &=& \sigma\big(\bar{\Psi}\Psi\to h^*\to X_{\textsc{sm}}\big) \,+\,
\sigma\big(\bar{\Psi}\Psi\to hh\big) \,,
\vphantom{|_{\int_|^|}^{}} \nonumber \\ \label{PP2h2sm}
\sigma\big(\bar{\Psi}\Psi\to h^*\to X_{\textsc{sm}}\big) &=&
\frac{\big(5\beta_{\Psi}^{}-6\beta_{\Psi}^3
+ 9\beta_{\Psi}^5\big)\lambda_{\Psi h\,}^2 s^{5/2}\,
\raisebox{3pt}{\footnotesize$\displaystyle\sum_i$}\,\Gamma\big(\tilde h\to X_{i,\textsc{sm}}\big)}
{576_{\,}m_{\Psi}^4\,
\big[\big(m_h^2-s\big)\raisebox{1pt}{$^2$}+\Gamma_h^2m_h^2\big]} \,, ~~~~~ ~~
X_{\textsc{sm}} \,\neq\, hh \,, ~~~~~
\end{eqnarray}
where $\beta_{\textsc x}$ is defined in Eq.\,(\ref{DD2h2sm}), the formula for
$\sigma(\bar{\Psi}\Psi\to hh)$ is described in Appendix\,\,\ref{formulas}, and the Higgs' width
\,$\Gamma_h^{}=\Gamma_h^{\textsc{sm}}+\Gamma(h\to\bar{\Psi}\Psi)$\, receives a contribution from
the invisible channel \,$h\to\bar{\Psi}\Psi$\, if \,$2m_{\Psi}^{}<m_h^{}$.\,
From Eq.\,(\ref{LP0}), we derive
\begin{eqnarray} \label{Gh2PP}
\Gamma\big(h\to\bar{\Psi}\Psi\big) \,=\, \frac{\lambda_{\Psi h\,}^2 m_h^{}}{8\pi}
\frac{\big(1-6{\texttt{\slshape R}}_{\Psi}^2
+ 18{\texttt{\slshape R}}_{\Psi}^4\big)}{9{\texttt{\slshape R}}_{\Psi}^4}
\big(1-4{\texttt{\slshape R}}_{\Psi}^2\big)^{3/2} \,, ~~~~~~~
{\texttt{\slshape R}}_{\Psi}^{} \,=\, \frac{m_{\Psi}^{}}{m_h^{}} \,,
\end{eqnarray}
which is subject to
\begin{eqnarray} \label{Bh2PP}
{\cal B}\big(h\to\bar{\Psi}\Psi\big) \,=\,
\frac{\Gamma\big(h\to\bar{\Psi}\Psi\big)}{\Gamma_h^{}} \,<\, 0.16
\end{eqnarray}
based on the aforementioned LHC Higgs data~\cite{atlas+cms}.
For the $h$-mediated scattering \,$\Psi N\to\Psi N$,\, the cross section is
\begin{eqnarray} \label{csel3/2}
\sigma_{\rm el}^N \,=\, \frac{\lambda_{\Psi h\,}^2g_{NNh\,}^2m_\Psi^2m_N^2}
{\pi\,\big(m_\Psi^{}+m_N^{}\big)\raisebox{1pt}{$^2$}m_h^4} \,.
\end{eqnarray}

\begin{figure}[h]
\includegraphics[width=73mm]{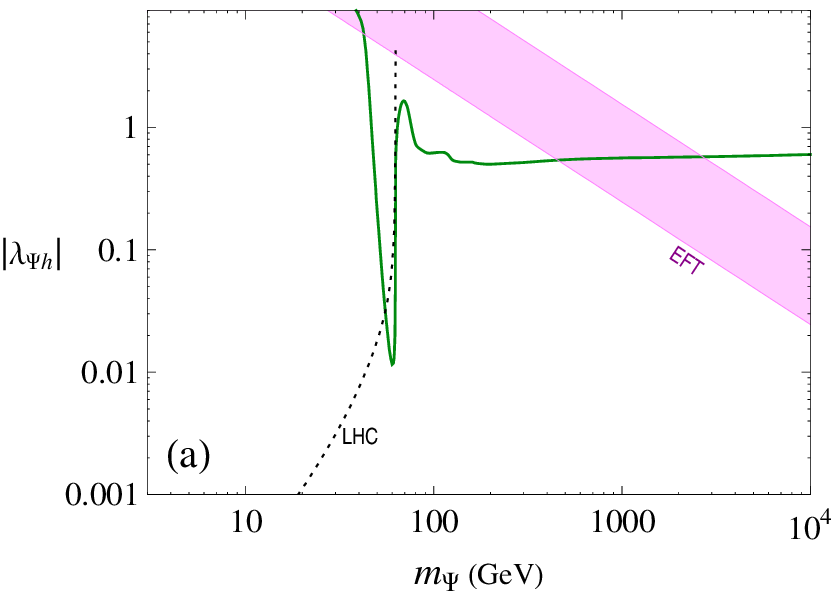} ~ ~
\includegraphics[width=75.3mm]{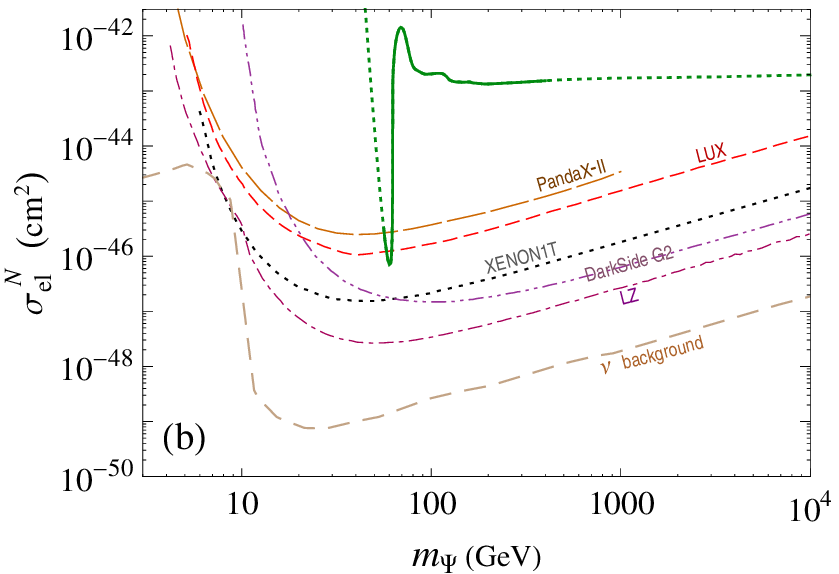}\vspace{-1ex}
\caption{The same as Fig.\,\ref{sm+fdm_half-plots}, except the DM is the spin-$3/2$
singlet $\Psi$ in the SM+$\Psi$.} \label{sm+Psi-plots}
\end{figure}

With the formulas above, we arrive at the $|\lambda_{\Psi h}|$ values consistent with
the observed relic density and the corresponding $\sigma_{\rm el}^N$, which are depicted by
the green curves in Figs.\,\,\ref{sm+Psi-plots}(a) and\,\,\ref{sm+Psi-plots}(b), respectively.
In Fig.\,\,\ref{sm+Psi-plots}(a), the black dotted curve marks the upper bound on
$|\lambda_{\Psi h}|$ from the Higgs measurements, and the magenta band represents
the zone \,$|\lambda_{\Psi h}|\in[1,2\pi]v/m_\Psi^{}$\, in which the EFT description may
be expected to have broken down, analogously to the spin-1/2 case.
Consequently, like before the lower boundary of this band may be taken to be the upper limit
for the validity of the EFT approximation.
The $m_\Psi$ range fulfilling these requirements translates into the solid part of
the green curve in Fig.\,\,\ref{sm+Psi-plots}(b) for the $\Psi$-nucleon cross-section,
$\sigma_{\rm el}^N$.
Evidently, the direct search bounds reduce the viable DM-mass region in the SM+$\Psi$
relative to the SM+$\psi$.
More precisely, only
\,58.0\,\,GeV\,$\mbox{\footnotesize$\lesssim$}\,m_\Psi\,
\mbox{\footnotesize$\lesssim$}\;$61.8\,\,GeV\,
remains viable.
This minuscule stretch toward the bottom of the Higgs-resonance dip is almost eliminated by
the final LUX limit and will likely be fully excluded by the forthcoming probes of PandaX-II,
unless it discovers $\Psi$.

\subsection{Vector dark matter\label{vdm}}

In the minimal model, dubbed SM+V, the only new ingredient beyond the SM is a spin-1 state
acting as the WIMP DM candidate~\cite{Kanemura:2010sh}.
We assume that it is described by a real field $\texttt{\slshape V}_\nu$ which is a singlet
under the SM gauge group and odd under an unbroken $Z_2$ symmetry which does not alter SM members.
Unlike its fermionic counterparts, $\texttt{\slshape V}_\nu$ can couple to the Higgs doublet
via a\,\,dimension-four operator.
The DM Lagrangian that respects the SM gauge symmetry is then~\cite{Kanemura:2010sh}
\begin{eqnarray}
{\cal L}_V^{} \,=\,
-\frac{1}{4} \texttt{\slshape V}_{\kappa\nu}^{}\texttt{\slshape V}^{\,\kappa\nu}
+ \frac{\mu_V^2}{2} \texttt{\slshape V}_\kappa^{}\texttt{\slshape V}^{\,\kappa}
+ \frac{\lambda_V^{}}{4}(\texttt{\slshape V}_\kappa^{}\texttt{\slshape V}^{\,\kappa})^2 +
\lambda_{h\,}^{}{\sf H}^\dagger{\sf H}\,\texttt{\slshape V}_\kappa^{}\texttt{\slshape V}^{\,\kappa} \,,
\end{eqnarray}
where
\,$\texttt{\slshape V}_{\kappa\nu} = \partial_\kappa \texttt{\slshape V}_\nu
- \partial_\nu \texttt{\slshape V}_\kappa$\,
and $\mu_V$, $\lambda_V$, and $\lambda_h$ are real constants.
Although the terms in ${\cal L}_V$ are at most of dimension four, it is actually
nonrenormalizable and violates unitarity~\cite{Lebedev:2011iq}.\footnote{This type of
Higgs-portal vector-DM model has been explored previously in~\cite{Kanemura:2010sh,
Djouadi:2011aa,Beniwal:2015sdl,Escudero:2016gzx,Kamenik:2011vy}.
Examples of its UV completion were proposed in \cite{Baek:2014jga,Hambye:2008bq}.\medskip}
Hereafter, we make no assumption about the details of the UV completion of ${\cal L}_V$,
implying that \texttt{\slshape V} is not necessarily a gauge boson and may even be
a composite object in the dark sector.
Consequently, we can generally treat $\mu_V$ and $\lambda_{V,h}$ as independent parameters.
For the calculations below, $\lambda_V$ does not play any essential role, and so we can express
\begin{eqnarray} \label{Lsm+v}
{\cal L}_V^{} \,\supset\,
\frac{m_V^2}{2} \texttt{\slshape V}_\kappa^{} \texttt{\slshape V}^{\,\kappa}
+ \lambda_h^{} \bigg(h v+\frac{h^2}{2}\bigg)
\texttt{\slshape V}_\kappa^{} \texttt{\slshape V}^{\,\kappa} \,,
\end{eqnarray}
where \,$m_V^{}=\big(\mu_V^2+\lambda_h v^2\big)\raisebox{1pt}{$^{1/2}$}$\, is
the \texttt{\slshape V} mass.

As in the fermionic DM models, $\lambda_h$ has to meet the different requirements on
DM annihilation, invisible decay \,$h\to\texttt{\slshape V}\texttt{\slshape V}$,\, and
DM-nucleon scattering.
From Eq.\,(\ref{Lsm+v}), we obtain
\begin{eqnarray}
\sigma_{\rm ann}^{} &=&
\sigma(\texttt{\slshape V}\texttt{\slshape V}\to h^*\to X_{\textsc{sm}}) \,+\,
\sigma(\texttt{\slshape V}\texttt{\slshape V}\to hh) \,,
\vphantom{|_{\int_|^|}^{}} \nonumber \\ \label{vv2h2sm}
\sigma(\texttt{\slshape V}\texttt{\slshape V}\to h^*\to X_{\textsc{sm}}) &=&
\frac{\lambda_h^2 \big(\beta_V^2s^2+12m_V^4\big) v^2~
\raisebox{3pt}{\footnotesize$\displaystyle\sum_i$}\,
\Gamma\big(\tilde h\to X_{i,\textsc{sm}}\big)}{9 \beta_{V\,}^{}m_V^4\sqrt s\;
\big[\big(m_h^2-s\big)\raisebox{1pt}{$^2$}+\Gamma_h^2m_h^2\big]}\,, ~~~~~ ~~
X_{\textsc{sm}} \,\neq\, hh \,, ~~~~~
\end{eqnarray}
the formula for $\sigma(\texttt{\slshape V}\texttt{\slshape V}\to hh)$ is relegated to
Appendix\,\,\ref{formulas} and the $h$ width
\,$\Gamma_h^{}=\Gamma_h^{\textsc{sm}}+\Gamma(h\to\texttt{\slshape V}\texttt{\slshape V})$,\,
with
\begin{eqnarray} \label{Gh2vv}
\Gamma(h\to\texttt{\slshape V}\texttt{\slshape V}) \,=\, \frac{\lambda_h^{2\,}v^2}{8\pi m_h^{}}
\frac{\big(1-4{\texttt{\slshape R}}_V^2+12{\texttt{\slshape R}}_V^4\big)}
{4 {\texttt{\slshape R}}_V^4} \sqrt{1-4{\texttt{\slshape R}}_V^2} \,, ~~~~~~~
{\texttt{\slshape R}}_V^{} \,=\, \frac{m_V^{}}{m_h^{}} \,.
\end{eqnarray}
We will again demand
\begin{eqnarray} \label{Bh2vv}
{\cal B}(h\to\texttt{\slshape V}\texttt{\slshape V}) \,=\,
\frac{\Gamma(h\to\texttt{\slshape V}\texttt{\slshape V})}{\Gamma_h^{}} \,<\, 0.16
\end{eqnarray}
based on the Higgs data~\cite{atlas+cms}.
For the Higgs-mediated DM-nucleon collision \,$\texttt{\slshape V}N\to\texttt{\slshape V}N$,\,
the cross section is
\begin{eqnarray} \label{vn2vn}
\sigma_{\rm el}^N \,=\, \frac{\lambda_{h\,}^2g_{NNh\,}^2m_N^2v^2}
{\pi\,\big(m_V^{}+m_N^{}\big)\raisebox{1pt}{$^2$}m_h^4} \,.
\end{eqnarray}

There are also theoretical considerations relevant to restraining $\lambda_h$.
Since ${\cal L}_V$ in Eq.\,(\ref{Lsm+v}) leads to unitarity violation~\cite{Lebedev:2011iq},
we need to ensure that it does not occur with the extracted $\lambda_h$ values.
As discussed in Appendix\,\,\ref{theory-reqs}, this implies that we need to have
\,$|\lambda_h^{}|<\sqrt{2\pi}\,m_V^{}/v$.\,
There is additionally a complementary restraint from the expectation that the theory
remains perturbative.
As also explained in Appendix\,\,\ref{theory-reqs}, this translates into the bound
\,$|\lambda_h^{}|<1$.\,

\begin{figure}[h]
\includegraphics[width=73mm]{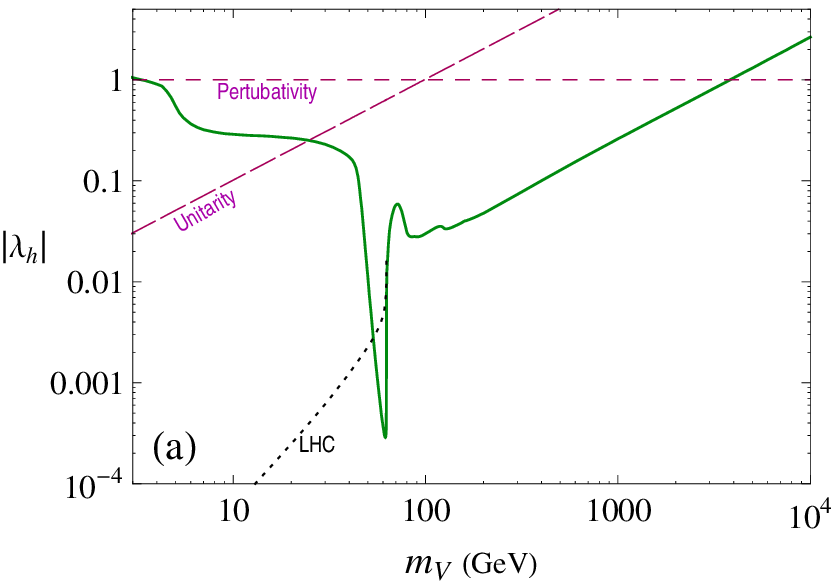} ~ ~
\includegraphics[width=75.3mm]{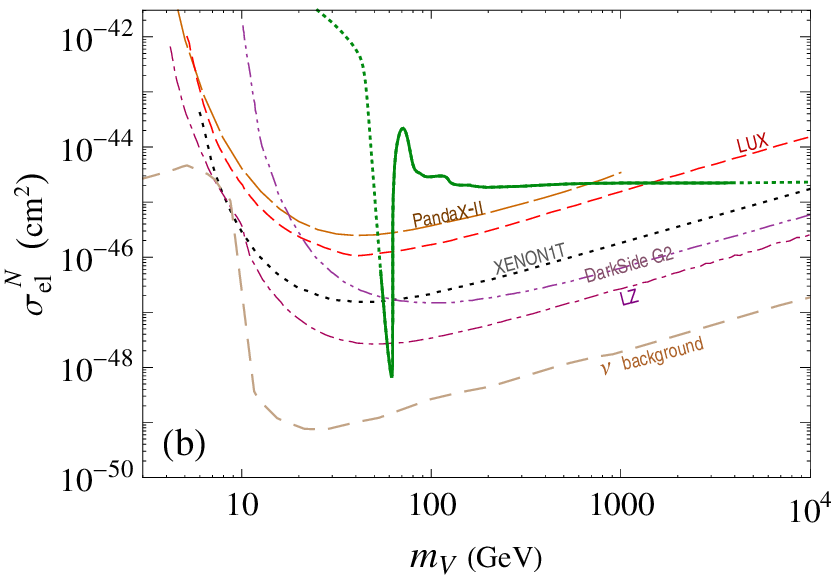}\vspace{-1ex}
\caption{The magnitude of Higgs coupling $\lambda_h$ to the vector DM versus its mass in
the SM+V satisfying the relic abundance requirement (green curve), compared to the upper
limits inferred from Higgs invisible decay data (black dotted curve) and from unitarity
and perturbativity considerations (maroon dashed curves).
(b)\,\,The corresponding cross-section $\sigma_{\rm el}^{N}$ of DM-nucleon scattering (green
curve) compared to the same data and projections as in Fig.\,\ref{sm+fdm_half-plots}(b).
The dotted sections of the green curve are disallowed by the constraints in (a).}
\label{sm+vdm-plots} \end{figure}

In Fig.\,\ref{sm+vdm-plots}(a), we present the $\lambda_h$ values fulfilling the relic density
requirement (green curve).
Also shown are the upper limits from the Higgs invisible decay data (black dashed curve) and
from the unitarity and perturbativity considerations (maroon dashed curves).
We plot the corresponding \texttt{\slshape V}-nucleon cross-section from Eq.\,(\ref{vn2vn})
in Fig.\,\ref{sm+vdm-plots}(b), where the dotted sections of the green curve are disallowed
by the restrictions in\,\,Fig.\,\ref{sm+vdm-plots}(a).
We find that $m_V$ values approximately below 54\,GeV and from 62.6\,GeV to 1.42 TeV
are in conflict with LHC and LUX data.
Comparable results were obtained in Ref.\,\cite{Escudero:2016gzx}.
However, the graphs also reveal that for \,$m_V^{}>3.9$\,TeV\,
the effective theory probably is no longer perturbative.
Thus, overall the situation is rather similar to that in the simplest scalar-DM
model~\cite{He:2016mls,Wu:2016mbe,Escudero:2016gzx}, where the annihilation rate
does not suffer from the $v_{\rm rel}^2$ suppression and consequently the viable
parameter space is far greater than in its fermionic counterparts.

\section{Two-Higgs-doublet-portal fermionic and vector DM models\label{sec:2hdm+fdm}}

Here we explore extensions of the minimal models in the last section by adding in each
case another Higgs doublet.
Furthermore, we suppose that the SM fermions in the extended scenarios have the Yukawa
interactions of the two-Higgs-doublet model (THDM) of type II, where the down-type fermions
get mass from only one of the Higgs doublets, $H_1$, and the up-type fermions from
the other doublet, $H_2$.
Accordingly, the Yukawa Lagrangian is~\cite{thdm,Branco:2011iw}
\begin{eqnarray} \label{LY2hdm}
{\cal L}_{\rm Y}^{} \,=\,
-\overline{Q}_{j,L}^{} \big(\lambda_2^u\big)_{jl} \tilde H_{2\,}^{}{\cal U}_{l,R}^{}
- \overline{Q}_{j,L}^{} \big(\lambda_1^d\big)_{jl} H_1^{} {\cal D}_{l,R}^{}
- \overline{L}_{j,L}^{} \big(\lambda_1^\ell\big)_{jl} H_1^{} E_{l,R}^{}
\;+\; {\rm H.c.} \,,
\end{eqnarray}
where \,$j,l=1,2,3$\, are summed over, $Q_{j,L}$ $(L_{j,L})$ stands for left-handed quark
(lepton) doublets, \,${\cal U}_{l,R}^{}$ and ${\cal D}_{l,R}$ $(E_{l,R})$ are right-handed quark
(charged lepton) fields, \,$\tilde H_{1,2}^{}=i\tau_2^{}H_{1,2}^*$\, with $\tau_2^{}$ being
the second Pauli matrix, and $\lambda^{u,d,\ell}$ represent 3$\times$3
Yukawa-coupling matrices.
This Lagrangian respects another discrete symmetry, $Z_2'$, under which \,$H_2\to-H_2$\, and
\,${\cal U}_R\to-{\cal U}_R$,\, while all the other fields are not affected.
Thus, $Z_2'$ prohibits the combinations \,$\overline{Q}{}_L\tilde{H}_1{\cal U}_R$,
$\overline{Q}{}_LH_2{\cal D}_R$, $\overline{L}{}_LH_2E_R$, and their Hermitian conjugates from
entering ${\cal L}_{\rm Y}$.

In the scalar sector, the renormalizable potential ${\cal V}_H$ is that of the THDM\,\,II,
\begin{eqnarray} \label{pot}
{\cal V}_H^{} &=& m_{11\,}^2 H_1^\dagger H_1^{} + m_{22\,}^2 H_2^\dagger H_2^{}
- \big(m_{12}^2\,H_1^\dagger H_2^{}\,+\,{\rm H.c.}\big) \,+\,
\frac{\lambda_1}{2} \bigl(H_1^\dagger H_1^{}\bigr)\raisebox{1pt}{$^2$} +
\frac{\lambda_2}{2} \bigl(H_2^\dagger H_2^{}\bigr)\raisebox{1pt}{$^2$} ~~~~~
\nonumber \\ && \! +\;
\lambda_{3\,}^{}H_1^\dagger H_{1\,}^{}H_2^\dagger H_2^{}
+ \lambda_{4\,}^{}H_1^\dagger H_{2\,}^{}H_2^\dagger H_1^{}
+ \frac{\lambda_5^{}}{2}
\Big[ \big(H_1^\dagger H_2^{}\big)\raisebox{1pt}{$^2$} \,+\,{\rm H.c.}\Big] \,.
\end{eqnarray}
Although dimension-4 combinations with an odd number of $H_2^{(\dagger)}$ cannot appear
due to $Z_2'$, in ${\cal V}_H$ we have allowed the $m_{12}^2$ terms which softly break $Z_2'$
and are important in relaxing the caps on the Higgs masses~\cite{Branco:2011iw}.
The Hermiticity of ${\cal V}_H$ implies that $m_{11,22}^2$ and $\lambda_{1,2,3,4}$ are real
parameters.
With ${\cal V}_H$ chosen to be $CP$ invariant, $m_{12}^2$ and $\lambda_5^{}$
are also real constants.

To see how ${\cal V}_H$ describes the couplings among the physical states in the Higgs
doublets, we first decompose them as
\begin{eqnarray}
H_r^{} \,=\, \frac{1}{\sqrt2} \left(\begin{array}{c} \sqrt2\,h_r^+ \vspace{1ex} \\
v_r^{}+ h_r^0 + i I_r^0 \end{array}\right ) \,, ~~~~~
r \,=\, 1,2 \,,
\end{eqnarray}
where $v_{1,2}^{}$ denote the VEVs of $H_{1,2}$, respectively, and are linked to
the electroweak mass scale \,$v\simeq246$\,GeV\, by \,$v_1^{}=v\cos\beta$\, and
\,$v_2^{}=v\sin\beta$.\,
The components $h_r^+$, $h_r^0$, and $I_r^0$ are connected to the physical Higgs
bosons $h$, $H$, $A$, and $H^+$ by
\begin{eqnarray}
\left(\begin{array}{c} h^+_1 \vspace{1ex} \\ h^+_2 \end{array}\right) &\!=\!&
\left(\begin{array}{lr} c_\beta^{}~ & -s_\beta^{} \vspace{3pt} \\
s_\beta^{} & c_\beta^{} \end{array}\right)
\left(\begin{array}{c} w^+ \vspace{1ex} \\ H^+ \end{array}\right) , ~~~~ ~~~
\left(\begin{array}{c} I_1^0 \vspace{1ex} \\ I_2^0 \end{array}\right) = \left(\begin{array}{lr}
c_\beta^{}~ & -s_\beta^{} \vspace{3pt} \\ s_\beta^{} & c_\beta^{} \end{array}\right)
\left(\begin{array}{c} z \vspace{3pt} \\ A \end{array}\right) ,
\nonumber \\
\left(\begin{array}{c} h_1^0 \vspace{1ex} \\ h_2^0 \end{array}\right) &\!=\!&
\left(\begin{array}{lr} c_\alpha^{}~ & -s_\alpha^{} \vspace{3pt} \\
s_\alpha^{} & c_\alpha^{} \end{array}\right)
\left(\begin{array}{c} H \vspace{3pt} \\ h \end{array}\right) , \hspace{4.7em}
c_{\cal X}^{} \,=\, \cos{\cal X} \,, ~~~~~ s_{\cal X}^{} \,=\, \sin{\cal X} \,,
\end{eqnarray}
where $\cal X$ is any angle or combination of angles and $w^\pm$ and $z$ are, respectively,
the would-be Goldstone bosons that will be eaten by the $W^\pm$ and $Z$ bosons.
We can then express the terms in ${\cal V}_H$ after electroweak symmetry breaking that are
relevant to our purposes as
\begin{eqnarray} \label{LH}
{\cal V}_H^{} &\supset&
\Big( \tfrac{1}{6} \lambda_{hhh}^{}h^2 + \tfrac{1}{2} \lambda_{hhH}^{}h H
+ \tfrac{1}{2} \lambda_{hHH}^{}H^2 + \tfrac{1}{2} \lambda_{hAA}^{}A^2
+ \lambda_{hH^+H^-}^{}H^+H^- \Big) h v
\nonumber \\ && \! +\;
\Big( \tfrac{1}{6} \lambda_{HHH}^{}H^2 + \tfrac{1}{2} \lambda_{HAA}^{}A^2
+ \lambda_{HH^+H^-}^{}H^+H^- \Big) H v \,,
\end{eqnarray}
where the $\lambda$s are linked to the physical Higgs masses~\cite{Branco:2011iw} and
the relations are listed in Ref.\,\cite{He:2016mls}.

Since $h$ and $H$ couple directly to the $W$ and $Z$ bosons, we need to take into account DM
annihilation into \,$W^+W^-$ and $ZZ$.\,
The pertinent interactions are given by
\begin{eqnarray} \label{vvh}
{\cal L} \,\supset\, \bigl(2m_W^2W^{+\nu}W_\nu^-+m_Z^2 Z^\nu Z_\nu^{}\bigr)
\bigg(k_V^h\,\frac{h}{v}+k_V^H\,\frac{H}{v}\bigg) \,, ~~~~ ~~~
k_V^h \,=\, s_{\beta-\alpha}^{} \,, ~~~~~ k_V^H \,=\, c_{\beta-\alpha}^{} \,. ~~~~~
\end{eqnarray}

The presence of the extra Higgs doublet not only offers another portal between the dark
and visible sectors, but also can produce modifications to the effective coupling between
a Higgs boson $\cal H$ and a nucleon $\cal N$ which is defined by
\begin{eqnarray}
{\cal L}_{\cal NNH}^{} \,=\, -g_{\cal NNH}^{}\, \overline{\cal N}{\cal N}{\cal H} \,, ~~~ ~~~~
{\cal N} \,=\, p,n \,, ~~~~~ {\cal H} \,=\, h,H \,,
\end{eqnarray}
and plays a crucial role in DM-nucleon collisions.
The potential changes spring from the quark-Higgs terms in Eq.\,(\ref{LY2hdm})
\begin{eqnarray}
{\cal L}_{\rm Y}^{} \,\supset\, -\raisebox{-7pt}{\Large$\stackrel{\sum}{\mbox{\scriptsize$q$}}$}\,
k_q^{\cal H}m_q^{}\,\overline{q}q\,\frac{\cal H}{v} \,, ~~~~ ~~~
k_{c,t}^{\cal H} \,=\, k_u^{\cal H} \,, ~~~~~ k_{s,b}^{\cal H} \,=\, k_d^{\cal H} \,, ~~~~
\end{eqnarray}
where the sum is over all quarks, \,$q=u,d,s,c,b,t$,\, and
\begin{eqnarray}  \label{kukdII}
k_u^h \,=\,  \frac{c_\alpha^{}}{s_\beta^{}} \,, ~~~~ ~~~
k_d^h \,=\, -\frac{s_\alpha^{}}{c_\beta^{}} \,, ~~~~ ~~~
k_u^H \,=\,  \frac{s_\alpha^{}}{s_\beta^{}} \,, ~~~~ ~~~
k_d^H \,=\,  \frac{c_\alpha^{}}{c_\beta^{}} \,,
\end{eqnarray}
which are generally different from the SM values \,$k_q^h=1$\, and \,$k_q^H=0$.\,
Relating the above quark- and hadronic-level quantities, one arrives numerically
at~\cite{He:2016mls,He:2008qm,He:2011gc}
\begin{eqnarray} \label{gNNH2hdm}
g_{pp\cal H}^{} \,=\, \big(0.5631\,k_u^{\cal H}+0.5599\,k_d^{\cal H}\big)\times10^{-3} \,, ~~~~~
g_{nn\cal H}^{} \,=\, \big(0.5481\,k_u^{\cal H}+0.5857\,k_d^{\cal H}\big)\times10^{-3} \,. ~~~
\end{eqnarray}
Setting \,$k_{u,d}^h=1$\, in the \,${\cal H}=h$\, formulas, we reproduce the SM values
\,$g_{pph,nnh}^{\textsc{sm}}\simeq0.0011$\, quoted in the last section.
However, if $k_{u,d}^{\cal H}$ are away from their SM expectations, $g_{pp\cal H}^{}$ and
$g_{nn\cal H}^{}$ can be very dissimilar, manifesting considerable isospin-violation.
Especially, if $k_{u,d}^{\cal H}$ have opposite signs, it may be possible to reduce
$g_{\cal NNH}^{}$ such that the DM effective interactions with nucleons become weak enough
to evade the experimental constraints.

This suggests that to address DM-nucleon collisions in models with a THDM-II portal it is
more appropriate to work with either the DM-proton or -neutron cross-section
($\sigma_{\rm el}^p$ or $\sigma_{\rm el}^n$, respectively) rather than the DM-nucleon
one ($\sigma_{\rm el}^N$) under the assumption of isospin conservation.
For comparison with experiment, in case the DM effective interactions with nucleons violate
isospin, the computed $\sigma_{\rm el}^{p,n}$ can be converted to
$\sigma_{\rm el}^N$ by means of~\cite{Kurylov:2003ra}
\begin{eqnarray} \label{ivdm}
\sigma_{\rm el}^N\,\raisebox{-8pt}{\Large$\stackrel{\sum}{\mbox{\scriptsize$i$}}$}\,
\eta_i^{}\,\mu_{A_i}^2A_i^2 \,=\,
\sigma_{\rm el}^p\,\raisebox{-8pt}{\Large$\stackrel{\sum}{\mbox{\scriptsize$i$}}$}\,\eta_i^{}\,
\mu_{A_i}^2\big[{\cal Z}+\big(A_i^{}-{\cal Z}\bigr)f_n^{}/f_p^{}\big]\raisebox{1pt}{$^2$} \,,
~~~~ ~~~ \sigma_{\rm el}^n \,=\, \sigma_{\rm el}^p\,f_n^2/f_p^2 \,,
\end{eqnarray}
where each sum is over isotopes of the element in the target material with which the DM
interacts dominantly, $\eta_i^{}$  $(A_i)$ stand for the fractional abundances (the nucleon
numbers) of the isotopes, \,$\mu_{A_i}=m_{A_i}m_\psi/\bigl(m_{A_i}+m_\psi\bigr)$,\,
with $m_{A_i}$ being the $i$th isotope's mass, $\cal Z$ represents the proton number of
the element, and $f_{p(n)}$ is the effective coupling of the DM to the proton (neutron).

In what follows, we select $h$ to be the 125-GeV Higgs boson.
Accordingly, $k_{d,u,V}^h$ in Eq.\,(\ref{kukdII}) need to be compatible with LHC
measurements on the $h$ couplings to SM fermions and electroweak bosons.
The modification to the \,$h\to X\bar X$\, interaction due to new physics can be
parameterized by $\kappa_X^{}$ in
\,$\kappa_X^2=\Gamma_{h\to X\bar X}^{}/\Gamma_{h\to X\bar X}^{\textsc{sm}}$.\,
Assuming that \,$|\kappa_{W,Z}|\le1$\, and the total width of $h$ can be altered by decay
modes beyond the SM, the ATLAS and CMS Collaborations have carried out
simultaneous fits to their Higgs data to extract~\cite{atlas+cms}
\begin{eqnarray} \label{kappaex}
\kappa_W^{} &=& 0.90\pm0.09 \,, ~~~ ~~~~ \kappa_t^{} \,=\, 1.43_{-0.22}^{+0.23} \,, ~~~ ~~~~
|\kappa_b^{}| \,=\, 0.57\pm0.16 \,, ~~~ ~~~~
|\kappa_\gamma| \,=\, 0.90_{-0.09}^{+0.10} \,, ~~~
\nonumber \\
\kappa_Z^{} &=& 1.00_{-0.08}^{} \,, ~~~~ ~~~~\, |\kappa_g| \,=\, 0.81_{-0.10}^{+0.13} \,,
~~~ ~~~\, |\kappa_\tau^{}| \,=\, 0.87_{-0.11}^{+0.12} \,,
\end{eqnarray}
where~\cite{atlas+cms}
\,$\kappa_\gamma^2=0.07\,\kappa_t^2+1.59\,\kappa_W^2-0.66\,\kappa_t^{}\kappa_W^{}$.\,
In the context of the THDM\,\,II, we expect these numbers to obey the relations
\,$k_V^h=\kappa_W^{}=\kappa_Z^{}$,\, $k_u^h=\kappa_t^{}\simeq\kappa_g^{}$,\, and
\,$k_d^h=\kappa_b^{}=\kappa_\tau^{}$\, within one sigma, but
the $\kappa_{t,g}^{}$ $(\kappa_{b,\tau})$ numbers above overlap only at the two-sigma level.
Pending improvement in the precision of these parameters from future data and following
Ref.\,\cite{He:2016mls}, based on Eq.\,(\ref{kappaex}) we can then impose
\begin{eqnarray} \label{khconstr}
0.81 \,\le\, k_V^h \,\le\, 1 \,, ~~~~~ 0.71 \,\le\, k_u^h \,\le\, 1.66 \,, ~~~~~
0.41 \,\le\, \big|k_d^h\big| \,\le\, 0.99 \,, ~~~~~
0.81 \,\le\, \big|k_\gamma^h\big| \,\le\, 1 \,, ~~~
\end{eqnarray}
where $k_\gamma^h$ includes the loop contribution of $H^\pm$ to \,$h\to\gamma\gamma$,\, and
so \,$k_\gamma^h\to\kappa_\gamma^{}$\, if the impact of $H^\pm$ is vanishing.

There are other constraints that we need to consider as well.
The extra Higgs particles in the THDM generally modify the so-called oblique electroweak
parameters $S$ and $T$ encoding the impact of new physics coupled to the standard SU(2)$_L$
gauge boson~\cite{Peskin:1991sw}, and so the new scalars must also conform to the empirical
requisites on these quantities.
To ensure this, we use the pertinent results of Ref.\,\cite{Grimus:2008nb} and
the $S$ and $T$ data from Ref.\,\cite{pdg}.

Theoretically, the parameters of the scalar potential \,${\cal V}_H$\, in Eq.\,(\ref{pot})
need to fulfill a\,\,number of conditions.
To keep the theory perturbative, each of the quartic couplings in ${\cal V}_H$ cannot be too big.
Another requirement is the stability of ${\cal V}_H$, implying that it has to be bounded
from below.
It is also essential to check that the (tree level) amplitudes for scalar-scalar
scattering do not violate unitarity constraints.
We summarize the expressions pertaining to these conditions in Appendix\,\,\ref{theory-reqs}.

In the rest of this section, we treat in turn the THDM\,\,II+$\psi$, THDM\,\,II+$\Psi$, and
THDM\,\,II+V, which are respectively the type-II two-Higgs-doublet extensions of
the minimal models of the previous section.
Thus, the DM stability is maintained in each case by the exactly conserved $Z_2$ symmetry as
before, under which the DM is odd and all the other fields are even.
In addition, we demand that, besides the scalar potential, the DM sector be $CP$ invariant.
We will demonstrate that in the presence of the second doublet it is possible to have
substantial weakening of the constraints from DM direct detection experiments or perhaps
even to evade them in the future.

\subsection{THDM II+\boldmath$\psi$\label{2hdm+psi}}

In this scenario, the Lagrangian for the DM is~\cite{Cai:2013zga,Banik:2013zla}
\begin{eqnarray} \label{Lpsi'}
{\cal L}_\psi' \,=\, \overline{\psi}\,i\slash{\!\!\!\partial}\psi \,-\,
\mu_\psi^{}\overline{\psi}_{\,\!}\psi \,-\, \overline{\psi}\psi\, \Bigg(
\frac{H_1^\dagger H_1^{}}{\Lambda_{1\psi}^{}}
+ \frac{H_2^\dagger H_2^{}}{\Lambda_{2\psi}^{}} \Bigg) \,,
\end{eqnarray}
where $\mu_\psi$ and $\Lambda_{1\psi,2\psi}$ are real constants of dimension mass and
$\Lambda_{1\psi,2\psi}$ contain the parameters of the underlying heavy new physics.
The $Z_2'$ symmetry prevents the combinations \,$\overline{\psi}\psi H_1^\dagger H_2^{}$\,
and \,$\overline{\psi}\psi H_2^\dagger H_1^{}$\, from appearing in ${\cal L}_\psi'$.

After electroweak symmetry breaking, we can express the relevant terms in
${\cal L}_\psi'$ involving the physical bosons as
\begin{eqnarray} \label{LHD}
{\cal L}_\psi' &\supset& -m_\psi^{}\overline{\psi}_{\,\!}\psi \,-\,
\overline{\psi}_{\,\!}\psi_{\,} \big(\lambda_{\psi h}^{} h + \lambda_{\psi H}^{} H\big)
\nonumber \\ && \! -\;
\frac{\overline{\psi}_{\,\!}\psi}{2 v} \big( \lambda_{\psi hh}^{}h^2
+ 2\lambda_{\psi hH\,}^{}h H + \lambda_{\psi HH}^{}H^2 + \lambda_{\psi AA}^{}A^2
+ 2\lambda_{\psi H^+H^-}^{}H^+H^- \big) \,,
\end{eqnarray}
where
\begin{eqnarray} \label{lambdah}
m_\psi &=& \mu_\psi
+ \big(\lambda_{1\psi\,}^{}c_\beta^2+\lambda_{2\psi\,}^{}s_\beta^2\big)\frac{v}{2} \,,
\hspace{9ex} \lambda_{r\psi}^{} \,=\, \frac{v}{\Lambda_{r\psi}} \,, ~~~~~ r \,=\, 1,2 \,,
\nonumber \\
\lambda_{\psi h}^{} &=& \lambda_{2\psi\,}^{}c_\alpha^{}s_\beta^{}
- \lambda_{1\psi\,}^{}s_\alpha^{}c_\beta^{} \,, \hspace{5.8em}
\lambda_{\psi H}^{} \,=\,
\lambda_{1\psi\,}^{}c_\alpha^{}c_\beta^{} + \lambda_{2\psi\,}^{}s_\alpha^{}s_\beta^{} \,,
\nonumber \\
\lambda_{\psi hh}^{} &=& \lambda_{1\psi\,}^{}s_\alpha^2
+ \lambda_{2\psi\,}^{}c_\alpha^2 \,, \hspace{7.2em}
\lambda_{\psi HH}^{} \,=\, \lambda_{1\psi\,}^{}c_\alpha^2 + \lambda_{2\psi\,}^{}s_\alpha^2 \,,
\nonumber \\
\lambda_{\psi hH}^{} &=& \big(\lambda_{2\psi}^{}-\lambda_{1\psi}^{}\big)c_\alpha^{}s_\alpha^{} \,,
\hspace{6.7em} \lambda_{\psi AA}^{} \,=\, \lambda_{\psi H^+H^-}^{} \,=\,
\lambda_{1\psi\,}^{}s_\beta^2 \,+\, \lambda_{2\psi\,}^{}c_\beta^2 \,. ~~~~~
\end{eqnarray}
There is no $\bar\psi\psi A$ term under the assumed $CP$ conservation.
Since $\mu_\psi$ and $\Lambda_{1\psi,2\psi}$ are free parameters, so are $m_\psi$
and~$\lambda_{\psi h,\psi H}$.
The couplings of $\bar\psi\psi$ to a pair of Higgs bosons can then be expressed
in terms of $\lambda_{\psi h,\psi H}$ as
\begin{eqnarray} \label{lambdahh}
\lambda_{\psi hh}^{} &=&\!
\Bigg(\frac{c_\alpha^3}{s_\beta^{}}-\frac{s_\alpha^3}{c_\beta^{}}\Bigg)\lambda_{\psi h}^{}
+ \frac{s_{2\alpha}^{}c_{\beta-\alpha}^{}}{s_{2\beta}^{}}\, \lambda_{\psi H}^{} \,, ~~~~~
\lambda_{\psi hH}^{} \,=\,
\frac{s_{2\alpha}^{}}{s_{2\beta}^{}}\big(\lambda_{\psi h\,}^{}c_{\beta-\alpha}^{}
- \lambda_{\psi H\,}^{}s_{\beta-\alpha}^{} \big) \,,
\nonumber \\
\lambda_{\psi HH}^{} &=&
\Bigg( \frac{c_\alpha^3}{c_\beta^{}}+\frac{s_\alpha^3}{s_\beta^{}} \Bigg) \lambda_{\psi H}^{}
- \frac{s_{2\alpha}^{}s_{\beta-\alpha}^{}}{s_{2\beta}^{}}\, \lambda_{\psi h}^{} \,,
\nonumber \\
\lambda_{\psi AA}^{} &=& \lambda_{\psi H^+H^-}^{} \,=\,
\frac{c_\alpha^{}c_\beta^3-s_\alpha^{}s_\beta^3}{c_\beta^{}s_\beta^{}}\, \lambda_{\psi h}^{}
+ \frac{c_\alpha^{}s_\beta^3+s_\alpha^{}c_\beta^3}{c_\beta^{}s_\beta^{}}\,
\lambda_{\psi H}^{} \,. ~~~~~
\end{eqnarray}

If both the $h$ and $H$ couplings to $\psi$ are nonzero, the DM-nucleon scattering
\,$\psi{\cal N}\to\psi\cal N$\, proceeds via tree-level diagrams mediated by $h$ and $H$,
leading to the cross section
\begin{eqnarray} \label{DN2DN}
\sigma_{\rm el}^{\cal N} \,=\,
\frac{m_\psi^2m_{\cal N}^2}{\pi\,\bigl(m_\psi+m_{\cal N}^{}\bigr)^2}
\Biggl( \frac{\lambda_{\psi h}^{}\,g_{{\cal NN}h}^{}}{m_h^2}
+ \frac{\lambda_{\psi H}^{}\,g_{{\cal NN}H}^{}}{m_H^2} \Biggr)^{\!\!2}
\end{eqnarray}
for momentum transfers small relative to $m_{h,H}$.
Given that the Higgs-nucleon coupling $g_{\cal NNH}^{}$, for \,${\cal N}=p$ or $n$\, and
\,${\cal H}=h$ or $H$,\, depends on $k_{u,d}^{\cal H}$ according to Eq.\,(\ref{gNNH2hdm}),
it may be possible to suppress $g_{\cal NNH}^{}$ sufficiently with a suitable choice of
$k_d^{\cal H}/k_u^{\cal H}$ to make $\sigma_{\rm el}^{\cal N}$ evade its experimental
limit~\cite{He:2008qm}, at least for some of the $m_\psi^{}$ values.
In addition, the $\lambda_{h,H}$ terms in Eq.\,(\ref{DN2DN}) may (partially) cancel each
other to lower $\sigma_{\rm el}^{\cal N}$ as well.
These are appealing features of the two-Higgs-doublet scenario that the one-doublet case
does not possess.
In evaluating model predictions for DM-nucleon reactions later on, we work exclusively
with the DM-proton cross-section, $\sigma_{\rm el}^p$, and then convert it to
$\sigma_{\rm el}^N$ with the aid of Eq.\,(\ref{ivdm}) for comparison with measurements.

As there are countless different possibilities in which $h$ and $H$ may act as portals
between the DM and other particles, for definiteness and simplicity hereafter we
concentrate on two scenarios in which the 125-GeV Higgs boson $h$ is lighter than
the other Higgs bosons, \,$m_h^{}<m_{H,A,H^\pm}$.\,
Moreover, we assume particularly that either $H$ or $h$ has a\,\,vanishing coupling to
the DM, \,$\lambda_{\psi H}=0$\, or \,$\lambda_{\psi h}=0$,\, respectively.
Accordingly, either $h$ or $H$ alone serves as the portal, and hence now we have
\,$f_n^{}/f_p^{}=g_{nn\cal H}^{}/g_{pp\cal H}^{}$,\, after ignoring the $n$-$p$ mass difference.

In the $h$-portal scenario ($\lambda_{\psi H}=0$), the cross section of DM annihilation is
\begin{eqnarray}
\sigma_{\rm ann}^{} \,=\, \sigma\big(\bar\psi\psi\to h^*\to X_{\textsc{sm}}\big) \,+\,
\raisebox{-7pt}{\Large$\stackrel{\sum}{\mbox{\scriptsize${\texttt s}_1{\texttt s}_2$}}$}\,
\sigma\big(\bar\psi\psi\to{\texttt s}_1^{}{\texttt s}_2^{}\big) \,,
\end{eqnarray}
where $\sigma\big(\bar\psi\psi\to h^*\to X_{\textsc{sm}}\big)$ is equal to that in
Eq.\,(\ref{DD2h2sm}), except the couplings of $h$ to fermions and gauge bosons are multiplied
by the appropriate $k_{u,d,V}^h$ factors mentioned earlier, the sum is over
\,${\texttt s}_1{\texttt s}_2=hh,hH,HH,AA,H^+H^-$\, with only kinematically permitted
channels contributing, and the formulas for
$\sigma\big(\bar\psi\psi\to{\texttt s}_1^{}{\texttt s}_2^{}\big)$
are listed in to Appendix\,\,\ref{formulas}.
After extracting $\lambda_{\psi h}$ from the relic density data and evaluating
$g_{pph}^{}$ with the $\alpha$ and $\beta$ choices consistent with Eq.\,(\ref{kukdII}),
we can predict the cross section of the $h$-mediated transition \,$\psi p\to\psi p$,\,
\begin{eqnarray} \label{Dp2Dp}
\sigma_{\rm el}^p \,=\, \frac{\lambda_{\psi h}^2\,g_{pph}^2\,m_\psi^2 m_p^2}
{\pi\,\bigl(m_\psi+m_p^{}\bigr)\raisebox{1pt}{$^2$}m_h^4} \,.
\end{eqnarray}
As in Sec.\,\ref{half}, for \,$2m_\psi^{}<m_h^{}$\, the invisible channel
\,$h\to\bar\psi\psi$\, is open, its rate already written down in Eq.\,(\ref{Gh2DD}).
Then ${\cal B}\big(h\to\bar\psi\psi\big)$ must be consistent with the LHC information on
the Higgs invisible decay, and so for this $m_\psi$ range we again impose
the bound in Eq.\,(\ref{lhclimit}).

In the $H$-portal case ($\lambda_{\psi h}=0$), the DM-annihilation cross-section is
\begin{eqnarray}
\sigma_{\rm ann}^{} \,=\, \sigma\big(\bar\psi\psi\to H^*\to X_{\textsc{sm}}\big) \,+\,
\raisebox{-9pt}{\Large$\stackrel{\sum}{\mbox{\scriptsize${\texttt s}_1{\texttt s}_2$}}$}\,
\sigma\big(\bar\psi\psi\to{\texttt s}_1^{}{\texttt s}_2^{}\big) \,,
\end{eqnarray}
where
\begin{eqnarray}
\sigma\big(\bar\psi\psi\to H^*\to X_{\textsc{sm}}\big) \,=\, \frac{\beta_{\psi\,}^{}
\lambda_{\psi H\,}^2\sqrt s~\raisebox{3pt}{\footnotesize$\displaystyle\sum_i$}\,
\Gamma\big(\tilde H\to X_{i,\textsc{sm}}\big)}
{2 \big[\big(m_H^2-s\big)\raisebox{1pt}{$^2$}+\Gamma_H^2m_H^2\big]} \,,
\end{eqnarray}
with $\tilde H$ being a virtual $H$ with mass \,$m_{\tilde H}=\sqrt s$.\,
Given that $H$ is not yet discovered, no empirical restraint on \,$H\to\bar\psi\psi$\, exists.
For the $\psi$-proton scattering via $H$ exchange, the cross section is
\begin{eqnarray} \label{Dp2Dp'}
\sigma_{\rm el}^p \,=\,
\frac{\lambda_{\psi H}^2\,g_{ppH}^2\,m_\psi^2 m_p^2}{\pi\,\bigl(m_\psi+m_p^{}\bigr)^2m_H^4} \,.
\end{eqnarray}
In applying Eq.\,(\ref{ivdm}), we set \,$f_n^{}/f_p^{}=g_{nnH}^{}/g_{ppH}^{}$.\,

\begin{table}[b] \footnotesize
\begin{tabular}{|c|cccccc|cccccccc|} \hline \scriptsize\,Set\, &
$\alpha$ & $\beta$ & $\displaystyle\frac{m_H^{}}{\scriptstyle\rm GeV}$ &
$\displaystyle\frac{m_A^{}}{\scriptstyle\rm GeV}$ &
$\displaystyle\frac{m_{H^\pm}^{}}{\scriptstyle\rm GeV}$ &
$\displaystyle\frac{m_{12}^2}{\scriptstyle\rm GeV^2}\,$ & $k_V^h$ & $k_u^h$ &
$\displaystyle\frac{k_d^h}{k_u^h}$ & $k_V^H$ & $k_u^H$ & $k_d^H$ &
\,$\displaystyle\frac{g_{pph}^{}}{10^{-5}}$\,
& $\displaystyle\frac{f_n^{}}{f_p}\vphantom{|_{\int_\int^\int}^{\int_\int^\int}}$
\\ \hline\hline
1 & \,0.141\,$\vphantom{\int^|}$ & \,1.422\, & \,550\, & 520 & 540 & 44000\, &
\,0.958\, & 1.001 & \,$-$0.947\, & \,0.286\, & 0.142 & \,6.68\, & \,3.29\, & \,$-$0.197\, \\
2 & 0.206$\vphantom{\int^|}$ & 1.357 & 515 & 560 & 570 &   55000\,   & \,0.913 & 1.002
& $-$0.962 & 0.408 & 0.209 & 4.61 & 2.42 & $-$0.646   \\
\hline \end{tabular}
\caption{Sample values of input parameters $\alpha$, $\beta$, $m_{H,A,H^\pm}$, and $m_{12}^2$
in the $h$-portal scenarios ($\lambda_{\psi H}=\lambda_{\Psi H}=\lambda_H=0$) and the resulting
values of several quantities, including \,$f_n^{}/f_p^{}=g_{nnh}^{}/g_{pph}^{}$.}
\label{lambdaH=0} \vspace{-3pt} \end{table}
\begin{table}[!b] \footnotesize
\begin{tabular}{|c|cccccc|cccccccc|} \hline \scriptsize\,Set\, &
$\alpha$ & $\beta$ & $\displaystyle\frac{m_H^{}}{\scriptstyle\rm GeV}$ &
$\displaystyle\frac{m_A^{}}{\scriptstyle\rm GeV}$ &
$\displaystyle\frac{m_{H^\pm}^{}}{\scriptstyle\rm GeV}$ &
$\displaystyle\frac{m_{12}^2}{\scriptstyle\rm GeV^2}\,$ & $k_V^h$ & $k_u^h$ & $k_d^h$ &
$k_V^H$ & $k_u^H$ & $\displaystyle\frac{k_d^H}{k_u^H}$ &
$\displaystyle\frac{g_{ppH}^{}}{10^{-5}}$
& $\displaystyle\frac{f_n^{}}{f_p}\vphantom{|_{\int_\int^\int}^{\int_\int^\int}}$
\\ \hline\hline
3 & \,$\vphantom{\int^|}-$0.749\, & \,$0.723$\, & \,610\, & 750 & 760 &   91000\, &
\,0.995\, & 1.107 & \,0.908\, & \,0.099\, & $-$1.029 & \,$-$0.949\, & \,$-$3.26\, & \,$-$0.245\, \\
4 & $\vphantom{\int^|}-$0.676 & $0.658$ & 590 & 610 & 640 &   60000\, &
  0.972   & 1.276 &   0.791 & 0.235 & $-$1.023 & $-$0.964 & $-$2.40 & $-$0.693 \\
\hline \end{tabular}
\caption{The same as Table \ref{lambdaH=0}, but for the $H$-portal scenarios
($\lambda_{\psi h}=\lambda_{\Psi h}=\lambda_h=0$).} \label{lambdah=0} \vspace{-9pt}
\end{table}
%%%

To illustrate the viable parameter space in these \,$\lambda_{\psi H(\psi h)}=0$\, scenarios,
we put together in Table\,\,\ref{lambdaH=0} (\ref{lambdah=0}) sample sets of input parameters
(the second to seventh columns) which are compatible with Eq.\,(\ref{khconstr}) and
the other requirements described in the two paragraphs following it.
The eighth to fifteenth columns of the tables contain the resulting values of several quantities.

\begin{figure}[b]
\begin{minipage}{0.45\textwidth}
\hspace{\fill}\includegraphics[width=75mm]{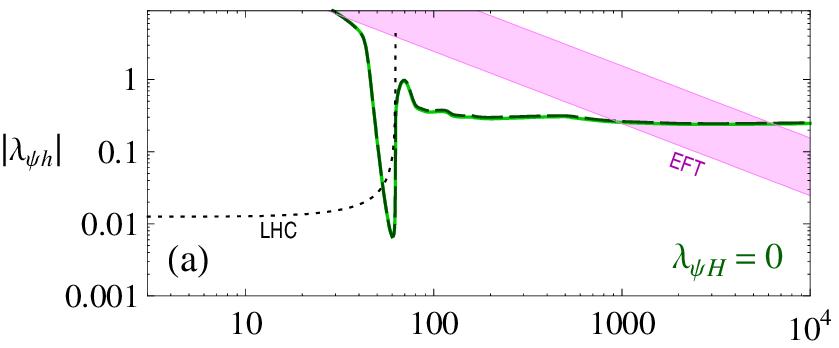}\vspace{-2ex}\\
\hspace{\fill}\includegraphics[width=76mm]{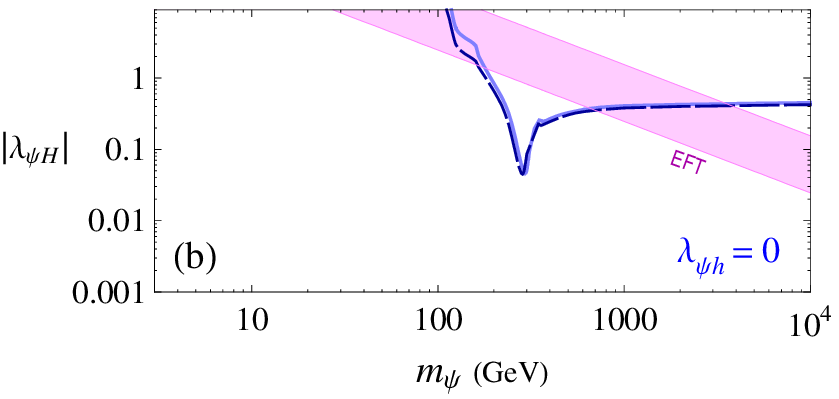}
\end{minipage}\begin{minipage}{0.55\textwidth}
\includegraphics[width=87mm]{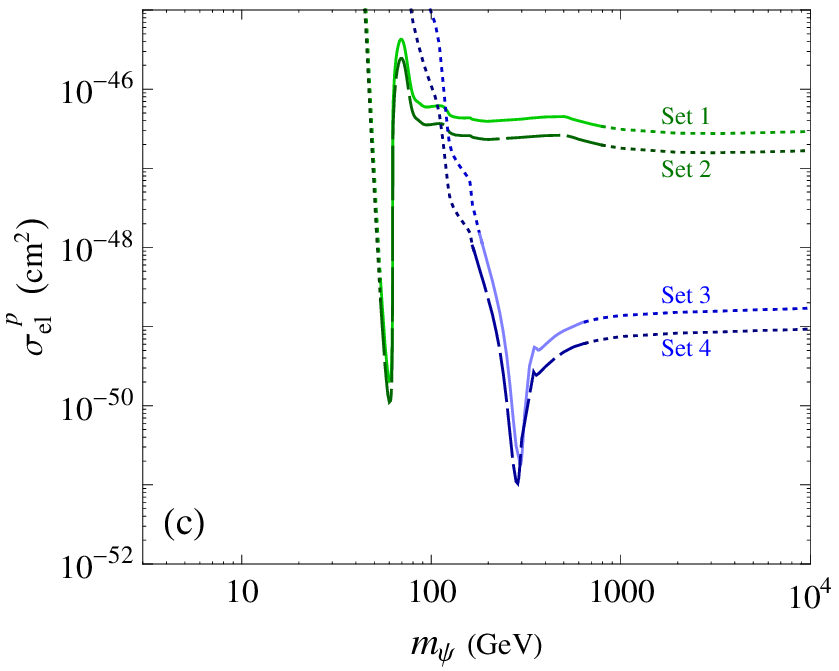}
\end{minipage} \vspace{-1ex}%
\caption{(a) The magnitude of DM-$h$ coupling $\lambda_{\psi h}$ consistent with the relic
density data versus $m_\psi$ in the THDM\,II+$\psi$ with \,$\lambda_{\psi H}=0$\, and input
numbers from Sets 1 (green solid curve) and 2 (green dashed curve) in Table\,\,\ref{lambdaH=0}.
Also plotted are upper limits inferred from LHC data on the Higgs invisible decay (black
dotted curve) and from the validity of the EFT approximation (lower sides of magenta bands).
(b)\,\,The same as (a), except with \,$\lambda_{\psi h}=0$\, and input numbers from
Sets 3 (blue solid curve) and 4 (blue dashed curve) in Table\,\,\ref{lambdah=0}, but without
the $h$ invisible decay restraint.
(c)\,\,The corresponding $\psi$-proton cross-sections $\sigma_{\rm el}^p$.
The dotted portions are disallowed by the constraints in (a) and (b).}
\label{2hdm+fdm_half_lambdas} \end{figure}

With these input numbers, we show in Figs.\,\,\ref{2hdm+fdm_half_lambdas}(a)
and\,\,\ref{2hdm+fdm_half_lambdas}(b) the $\lambda_{\psi h}$ and $\lambda_{\psi H}$ regions
evaluated from the observed relic density.
One observes that the $|\lambda_{\psi H}|$ values extracted from the relic density data
tend to be bigger than their $\lambda_{\psi h}$ counterparts.
This is because the $H$-portal annihilation rate is relatively more suppressed due to
\,$m_H^{}>m_h^{}$.\,
In Fig.\,\,\ref{2hdm+fdm_half_lambdas}(a), we also display the upper bound on
$|\lambda_{\psi h}|$ inferred from Eq.\,(\ref{lhclimit}) for the $h$ invisible decay (black
dotted curve).
Like in the minimal model of Sec.\,\ref{half}, the limited extent of the reliability of
the EFT approximation for the $\psi$-Higgs operators in ${\cal L}_\psi'$ implies that
in each $\cal H$-portal instance we also need to ensure
\,$|\lambda_{\psi\cal H}|<v/m_\psi$\,
beyond which the EFT framework may be expected to break down.
This condition is represented by the lower sides of the magenta bands in
Figs.\,\,\ref{2hdm+fdm_half_lambdas}(a) and\,\,\ref{2hdm+fdm_half_lambdas}(b).
We exhibit the corresponding predictions for $\sigma_{\rm el}^p$ in
Fig.\,\,\ref{2hdm+fdm_half_lambdas}(c), where the dotted parts of the green and blue curves
are excluded by the constraints in Figs.\,\,\ref{2hdm+fdm_half_lambdas}(a) and
\ref{2hdm+fdm_half_lambdas}(b), respectively.

\begin{figure}[t]
\includegraphics[width=85mm]{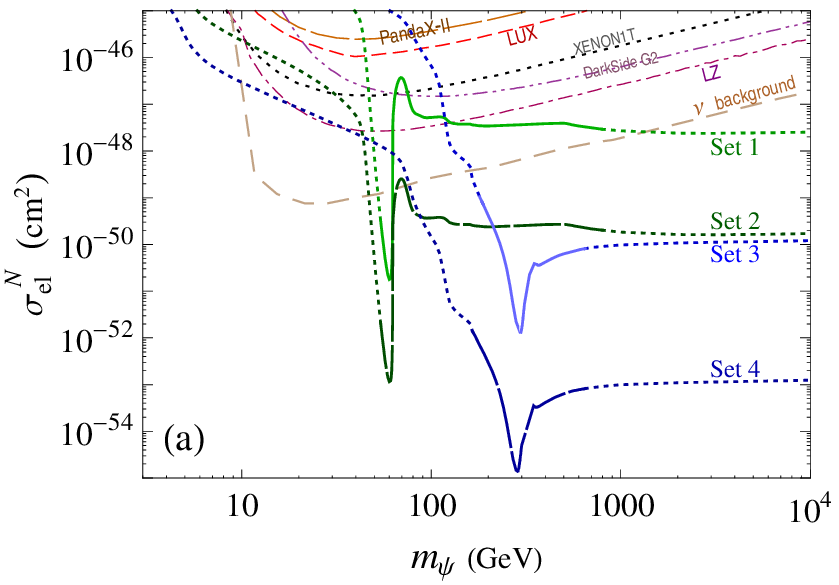}
\includegraphics[width=85mm]{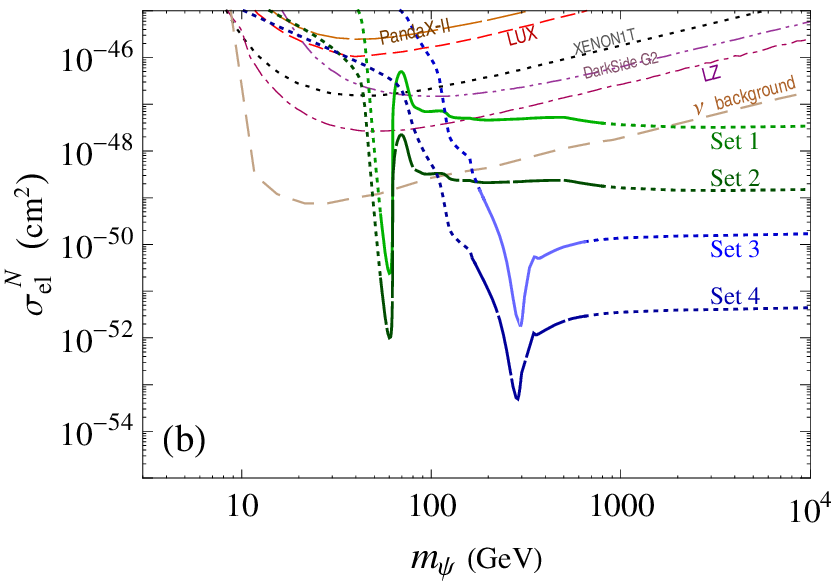}\vspace{-1ex}
\caption{The predicted DM-nucleon cross-sections in the THDM II+$\psi$ with input
numbers from Sets 1 and 2 (green curves) in Table\,\,\ref{lambdaH=0} and
Sets 3 and 4 (blue curves) in Tables\,\,\ref{lambdah=0} for (a)\,\,xenon and (b)\,\,argon targets,
compared to the same data and projections as in Fig.\,\,\ref{sm+fdm_half-plots}(b).
The dotted portions of the green and blue curves are excluded as in
Fig.\,\,\ref{2hdm+fdm_half_lambdas}.}
\label{2hdm+fdm_half_sigmaelN} \end{figure}

To test the model with direct search results, which are typically reported in terms of
the DM-nucleon cross-section $\sigma_{\rm el}^N$, we have converted the calculated
$\sigma_{\rm el}^p$ in Fig.\,\,\ref{2hdm+fdm_half_lambdas} to the (green and blue)
$\sigma_{\rm el}^N$ curves in Fig.\,\,\ref{2hdm+fdm_half_sigmaelN}(a) by means of
Eq.\,(\ref{ivdm}) with the $f_n^{}/f_p^{}$ values from the tables, assuming that
the target material in the detector is xenon.
Since the DarkSide G2 experiment will employ an argon target~\cite{dsg2}, in
Fig.\,\,\ref{2hdm+fdm_half_sigmaelN}(b) we plot the corresponding predictions for
$\sigma_{\rm el}^N$ assuming an argon target instead.
These graphs reveal some visible differences, especially the xenon curves for Sets 2 and 4
which are significantly lower than their argon counterparts.
The differences are not unexpected because the $f_n^{}/f_p^{}$ numbers in these
instances are not far from the xenophobic extreme, \,$f_n^{}/f_p^{}\simeq-0.7$.\,

Also depicted in Fig.\,\,\ref{2hdm+fdm_half_sigmaelN} are the same data and projections as
in Fig.\,\,\ref{sm+fdm_half-plots}(b).
It is obvious that, in stark contrast to the SM+$\psi$, the THDM\,\,II+$\psi$ accommodates a good
amount of parameter space which can evade the current direct search restrictions very well.
Particularly in these examples, over wide stretches of $m_\psi$ the model prediction
can also escape future direct detection and even hide below the neutrino floor.
In the $h$-portal cases, the LHC Higgs invisible decay data and the EFT validity limit
rule out \,$m_\psi<54$\,GeV\, and \,$m_\psi>0.9$\,TeV,\, respectively.
In the $H$-portal ones, the LHC Higgs invisible decay restraint does not apply,
but the EFT validity limit disallows $m_\psi$ values below 165 GeV and above 0.7 TeV.

\subsection{THDM II+\boldmath$\Psi$\label{2hdm+P}}

In this model, the Lagrangian for the spin-3/2 DM is
\begin{eqnarray}
{\cal L}_\Psi' \,=\, -\overline{\Psi_\nu}\big(i\slash{\!\!\!\partial}-\mu_\Psi^{}\big)\Psi^\nu
+ \overline{\Psi_\nu}_{\,}\Psi^\nu\, \Bigg(\frac{H_1^\dagger H_1^{}}{\Lambda_{1\Psi}^{}}
+ \frac{H_2^\dagger H_2^{}}{\Lambda_{2\Psi}^{}} \Bigg) \,,
\end{eqnarray}
where $\mu_\Psi$ and $\Lambda_{1\Psi,2\Psi}$ are
real constants of dimension mass and the latter two contain the parameters of the underlying
heavy new physics.
After electroweak symmetry breaking, we can then express the relevant terms in
${\cal L}_\Psi'$ involving the physical bosons as
\begin{eqnarray} \label{LPsi}
{\cal L}_\Psi' &\supset& m_\Psi^{}\,\overline{\Psi_\nu}_{\,}\Psi^\nu \,+\,
\overline{\Psi_\nu}_{\,}\Psi^\nu \big(\lambda_{\Psi h}^{} h + \lambda_{\Psi H}^{} H\big)
\nonumber \\ && \! +\;
\frac{\overline{\Psi_\nu}_{\,}\Psi^\nu}{2 v} \big( \lambda_{\Psi hh}^{}h^2
+ 2\lambda_{\Psi hH\,}^{}h H + \lambda_{\Psi HH}^{}H^2 + \lambda_{\Psi AA}^{}A^2
+ 2\lambda_{\Psi H^+H^-}^{}H^+H^- \big) \,,
\end{eqnarray}
where $m_\Psi$ and the $\lambda$s are the same in form as their counterparts
in Eqs.\,\,(\ref{lambdah}) and (\ref{lambdahh}), but with $\psi$ in the subscripts
replaced by $\Psi$.

It follows that the DM-annihilation cross-section in the $h$-portal scenario is
\begin{eqnarray}
\sigma_{\rm ann}^{} &=&
\sigma\big(\bar{\Psi}\Psi\to h^*\to X_{\textsc{sm}}\big) \,+\,
\raisebox{-7pt}{\Large$\stackrel{\sum}{\mbox{\scriptsize${\texttt s}_1{\texttt s}_2$}}$}\,
\sigma\big(\bar{\Psi}\Psi\to{\texttt s}_1^{}{\texttt s}_2^{}\big) \,,
\end{eqnarray}
where $\sigma\big(\bar{\Psi}\Psi\to h^*\to X_{\textsc{sm}}\big)$ is
equal to that in Eq.\,(\ref{PP2h2sm}), except the $h$ couplings of to SM particles are scaled
by the suitable $k_{u,d,V}^h$ factors, the sum is again over
\,${\texttt s}_1{\texttt s}_2=hh,hH,HH,AA,H^+H^-$,\, and the formulas for
$\sigma\big(\bar{\Psi}\Psi\to{\texttt s}_1^{}{\texttt s}_2^{}\big)$
are collected in Appendix\,\,\ref{formulas}.
As in Sec.\,\ref{rs}, for \,$2m_\Psi^{}<m_h^{}$\, the invisible channel
\,$h\to\bar{\Psi}\Psi$\, is open, its rate given by
Eq.\,(\ref{Gh2PP}), and so it must fulfill the condition in Eq.\,(\ref{Bh2PP}).
The $\Psi$-proton cross-section is the same as that in Eq.\,(\ref{Dp2Dp})
but with $\psi$ in the subscripts replaced by $\Psi$.

In the $H$-portal scenario
\begin{eqnarray}
\sigma_{\rm ann}^{} &=&
\sigma\big(\bar{\Psi}\Psi\to H^*\to X_{\textsc{sm}}\big) \,+\,
\raisebox{-7pt}{\Large$\stackrel{\sum}{\mbox{\scriptsize${\texttt s}_1{\texttt s}_2$}}$}\,
\sigma\big(\bar{\Psi}\Psi\to{\texttt s}_1^{}{\texttt s}_2^{}\big) \,,
\end{eqnarray}
where
\begin{eqnarray}
\sigma\big(\bar{\Psi}\Psi\to H^*\to X_{\textsc{sm}}\big) \,=\,
\frac{\big(5\beta_\Psi^{}-6\beta_\Psi^3
+ 9\beta_\Psi^5\big)\lambda_{\Psi H\,}^2 s^{5/2}\,
\raisebox{3pt}{\footnotesize$\displaystyle\sum_i$}\,\Gamma\big(\tilde H\to X_{i,\textsc{sm}}\big)}
{576_{\,}m_\Psi^4\,
\big[\big(m_H^2-s\big)\raisebox{1pt}{$^2$}+\Gamma_H^2m_H^2\big]} \,.
\end{eqnarray}
The cross section of $H$-mediated $\Psi$-proton scattering is equal to that in
Eq.\,(\ref{Dp2Dp'}), but with $\psi$ in the subscripts replaced by $\Psi$.

\begin{figure}[b]
\begin{minipage}{0.45\textwidth}
\hspace{\fill}\includegraphics[width=75mm]{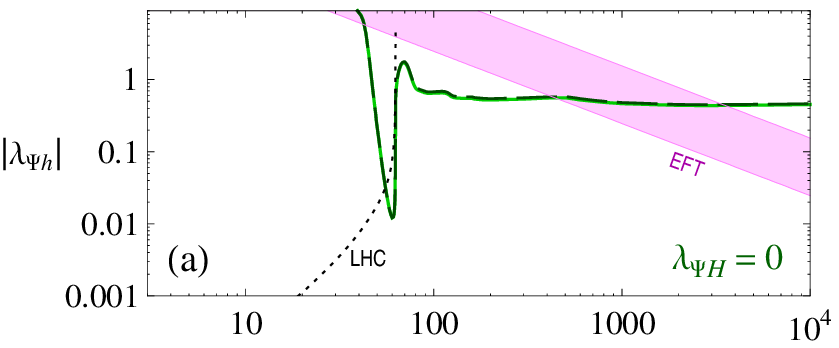}\vspace{-2ex}\\
\hspace{\fill}\includegraphics[width=76mm]{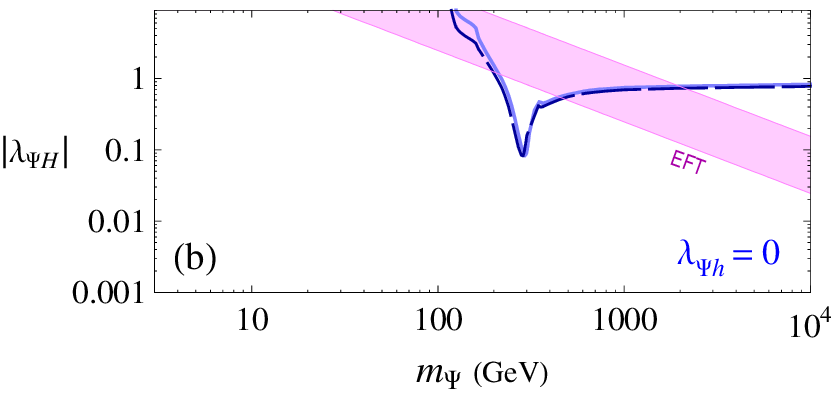}
\end{minipage}\begin{minipage}{0.55\textwidth}
\includegraphics[width=87mm]{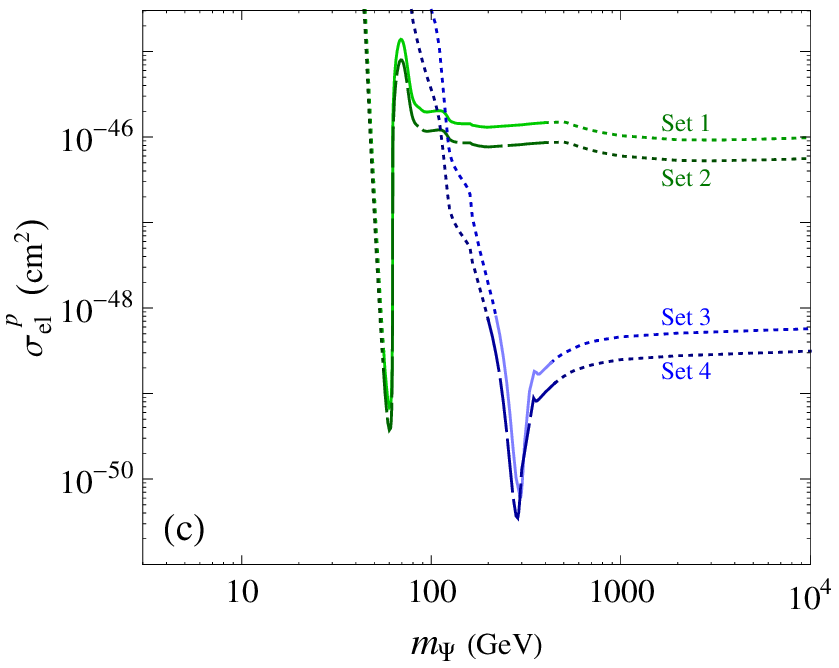}
\end{minipage} \vspace{-1ex}%
\caption{The same as Fig.\,\ref{2hdm+fdm_half_lambdas}, except the DM is the spin-$3/2$
singlet $\Psi$ in the THDM II+$\Psi$.} \label{2hdm+Psi_lambdas}
\end{figure}
\begin{figure}[t]
\includegraphics[width=85mm]{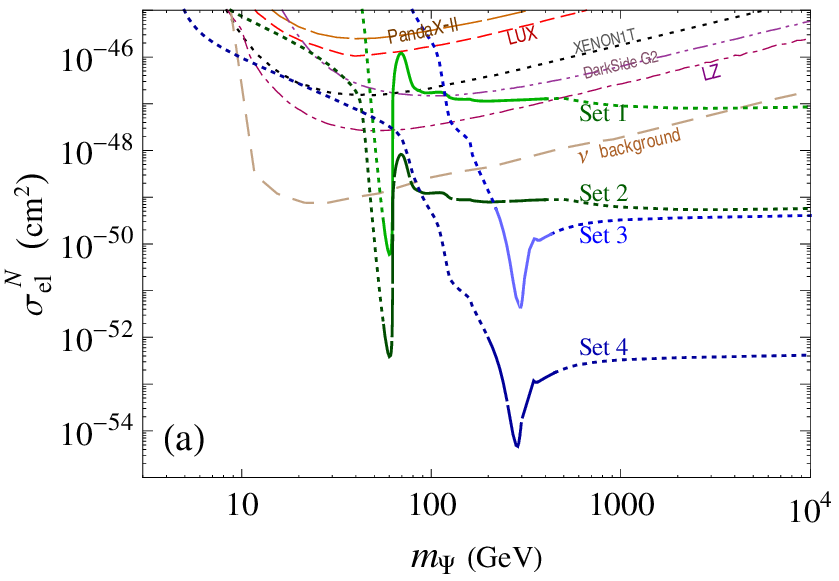}
\includegraphics[width=85mm]{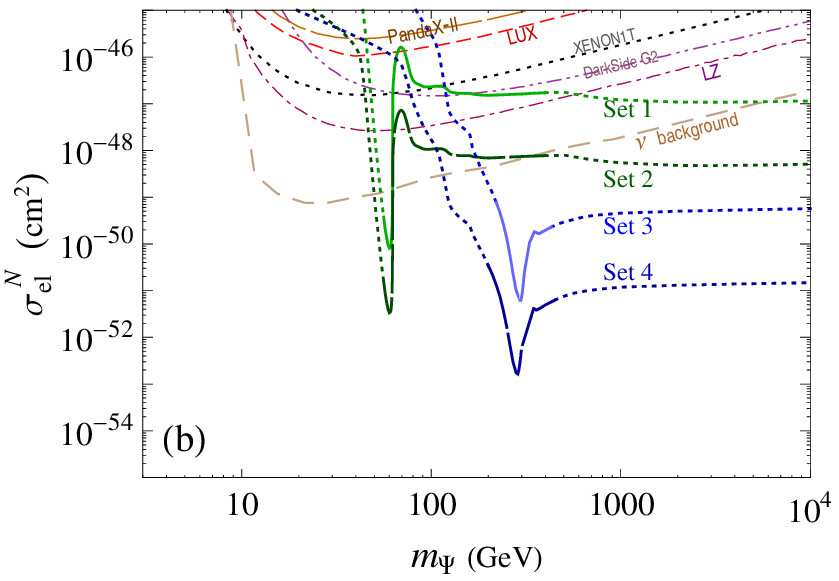}\vspace{-1ex}
\caption{The same as Fig.\,\ref{2hdm+fdm_half_sigmaelN}, except the DM is the spin-$3/2$
singlet $\Psi$ in the THDM II+$\Psi$.} \label{2hdm+Psi_sigmaelN}
\end{figure}

Similarly to the THDM II+$\psi$, we use the numbers from Tables\,\,\ref{lambdaH=0}
and\,\,\ref{lambdah=0} for our examples.
We present the results in Figs.\,\,\ref{2hdm+Psi_lambdas} and\,\,\ref{2hdm+Psi_sigmaelN}.
These instances indicate that in the THDM II+$\Psi$ the situation is roughly similar to
that in the THDM II+$\psi$, but the parameter space in the former is less able than in
the latter to escape the different restrictions.
Specifically for these examples,
we estimate the viable zones in the $h$- and $H$-portal cases, respectively, to be
\,$56{\rm\;GeV}\mbox{\footnotesize\,$\lesssim$\,}m_\Psi^{}\mbox{\footnotesize\,$\lesssim$\,}
420$\;GeV\, and
\,$200{\rm\;GeV}\mbox{\footnotesize\,$\lesssim$\,}m_\Psi^{}\mbox{\footnotesize\,$\lesssim$\,}
450$\;GeV,\,
which translate into predictions for $\sigma_{\rm el}^N$ which are currently below its
experimental limits.

\subsection{THDM II+V\label{2hdm+v}}

In this scenario, the Lagrangian for the vector DM is~\cite{Cai:2013zga}
\begin{eqnarray} \label{L2hdmv}
{\cal L}_V' &=&
-\frac{1}{4} \texttt{\slshape V}_{\kappa\nu}^{}\texttt{\slshape V}^{\,\kappa\nu}
+ \frac{\mu_V^2}{2} \texttt{\slshape V}_\kappa^{}\texttt{\slshape V}^{\,\kappa}
+ \frac{\lambda_V^{}}{4}(\texttt{\slshape V}_\kappa^{}\texttt{\slshape V}^{\,\kappa})^2
+ \big(  \lambda_{1\tt V}^{}H_1^\dagger H_1^{}+\lambda_{2\tt V}^{}H_2^\dagger H_2^{} \big)
\texttt{\slshape V}_\kappa^{}\texttt{\slshape V}^{\,\kappa}
\nonumber \\ &\supset&
\frac{m_V^2}{2} \texttt{\slshape V}_\kappa^{}\texttt{\slshape V}^{\,\kappa} +
\big(\lambda_h^{}h+\lambda_H^{}H\big)\texttt{\slshape V}_\kappa^{}\texttt{\slshape V}^{\,\kappa}v
\nonumber \\ && \! +\;
\frac{1}{2}\big( \lambda_{hh}^{}h^2+2\lambda_{hH}^{}h H+\lambda_{HH}^{}H^2+\lambda_{AA}^{}A^2 +
2\lambda_{H^+H^-}^{}H^+H^- \big) \texttt{\slshape V}_\kappa^{}\texttt{\slshape V}^{\,\kappa} \,,
\end{eqnarray}
where, analogously to the fermionic models,
\begin{eqnarray}
m_V^2 &=& \mu_V^2 + \bigl(\lambda_{1\tt V\,}^{}c_\beta^2
+ \lambda_{2\tt V\,}^{} s_\beta^2\bigr)v^2 \,, ~~~~
\lambda_h^{} \,=\, \lambda_{2\tt V\,}^{} c_\alpha^{}s_\beta^{}
- \lambda_{1\tt V\,}^{} s_\alpha^{}c_\beta^{} \,, ~~~~
\lambda_H^{} \,=\, \lambda_{1\tt V\,}^{} c_\alpha^{}c_\beta^{}
+ \lambda_{2\tt V\,}^{} s_\alpha^{}s_\beta^{} \,,
\nonumber \\
\lambda_{hh}^{} &=&
\Bigg(\frac{c_\alpha^3}{s_\beta^{}}-\frac{s_\alpha^3}{c_\beta^{}}\Bigg)\lambda_h^{}
+ \frac{s_{2\alpha}^{}c_{\beta-\alpha}^{}}{s_{2\beta}^{}}\,\lambda_H^{} \,, ~~~~ ~~~~
\lambda_{HH}^{} \,=\,
\Bigg( \frac{c_\alpha^3}{c_\beta^{}}+\frac{s_\alpha^3}{s_\beta^{}} \Bigg) \lambda_H^{}
- \frac{s_{2\alpha}^{}s_{\beta-\alpha}^{}}{s_{2\beta}^{}}\,\lambda_h^{} \,,
\nonumber \\
\lambda_{hH}^{} &=&
\frac{s_{2\alpha}^{}}{s_{2\beta}^{}}\big(\lambda_h^{}c_{\beta-\alpha}^{}
- \lambda_H^{}s_{\beta-\alpha}^{} \big) \,, ~~~~ ~~~~\,
\lambda_{AA,H^+H^-}^{} \,=\,
\frac{c_\alpha^{}c_\beta^3-s_\alpha^{}s_\beta^3}{c_\beta^{}s_\beta^{}}\lambda_h^{}
+ \frac{c_\alpha^{}s_\beta^3+s_\alpha^{}c_\beta^3}{c_\beta^{}s_\beta^{}}\lambda_H^{} \,.
\end{eqnarray}
We again look at separate possibilities with either \,$\lambda_H=0$\, or \,$\lambda_h=0$.\,

Thus, the cross section of the $h$-portal DM-annihilation is
\begin{eqnarray}
\sigma_{\rm ann}^{} &=&
\sigma(\texttt{\slshape V}\texttt{\slshape V}\to h^*\to X_{\textsc{sm}}) \,+\,
\raisebox{-9pt}{\Large$\stackrel{\sum}{\mbox{\scriptsize${\texttt s}_1{\texttt s}_2$}}$}\,
\sigma(\texttt{\slshape V}\texttt{\slshape V}\to{\texttt s}_1^{}{\texttt s}_2^{}) \,,
\end{eqnarray}
where $\sigma(\texttt{\slshape V}\texttt{\slshape V}\to h^*\to X_{\textsc{sm}})$ equals
that in Eq.\,(\ref{vv2h2sm}), but with the $h$ couplings to SM members
multiplied by the proper $k_{u,d,V}^h$ factors.
The expressions for
$\sigma(\texttt{\slshape V}\texttt{\slshape V}\to{\texttt s}_1^{}{\texttt s}_2^{})$
can be found in Appendix\,\,\ref{formulas}.
The invisible channel \,$h\to\texttt{\slshape V}\texttt{\slshape V}$\, has the rate already
given in Eq.\,(\ref{Gh2vv}) and hence is also constrained by Eq.\,(\ref{Bh2vv}).
For the $h$-mediated DM-nucleon collision \,$\texttt{\slshape V}N\to\texttt{\slshape V}N$,\,
the cross section is
\begin{eqnarray} \label{hportal}
\sigma_{\rm el}^p \,=\, \frac{\lambda_{h\,}^2g_{pph\,}^2m_p^2v^2}
{\pi\,\big(m_V^{}+m_p^{}\big)\raisebox{1pt}{$^2$}m_h^4} \,.
\end{eqnarray}

In the $H$-portal scenario, the cross section of DM annihilation is
\begin{eqnarray}
\sigma_{\rm ann}^{} \,=\,
\sigma(\texttt{\slshape V}\texttt{\slshape V}\to H^*\to X_{\textsc{sm}}) \,+\,
\raisebox{-9pt}{\Large$\stackrel{\sum}{\mbox{\scriptsize${\texttt s}_1{\texttt s}_2$}}$}\,
\sigma(\texttt{\slshape V}\texttt{\slshape V}\to{\texttt s}_1^{}{\texttt s}_2^{}) \,,
\end{eqnarray}
where
\begin{eqnarray}
\sigma(\texttt{\slshape V}\texttt{\slshape V}\to H^*\to X_{\textsc{sm}}) \,=\,
\frac{\lambda_H^2 \big(\beta_V^2s^2+12m_V^4\big) v^2~
\raisebox{3pt}{\footnotesize$\displaystyle\sum_i$}\,
\Gamma\big(\tilde H\to X_{i,\textsc{sm}}\big)}{9 \beta_{V\,}^{}m_V^4\sqrt s\;
\big[\big(m_H^2-s\big)\raisebox{1pt}{$^2$}+\Gamma_H^2m_H^2\big]}\,.
\end{eqnarray}
The cross section of the $H$-mediated \,$\texttt{\slshape V}N\to\texttt{\slshape V}N$\,
scattering has the formula in Eq.\,(\ref{hportal}), but with $h$ in the subscripts
replaced by $H$.

\begin{figure}[b]
\begin{minipage}{0.45\textwidth}
\hspace{\fill}\includegraphics[width=75mm]{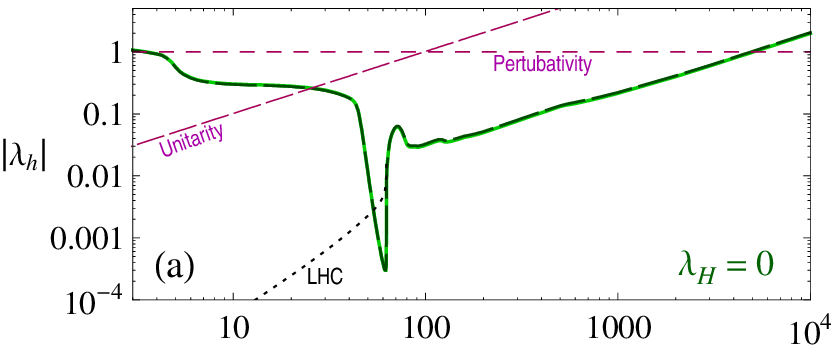}\vspace{-2ex}\\
\hspace{\fill}\includegraphics[width=76mm]{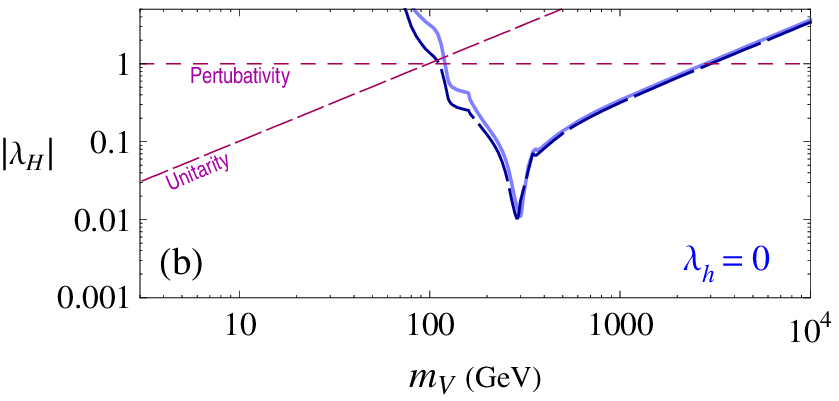}
\end{minipage}\begin{minipage}{0.55\textwidth}
\includegraphics[width=87mm]{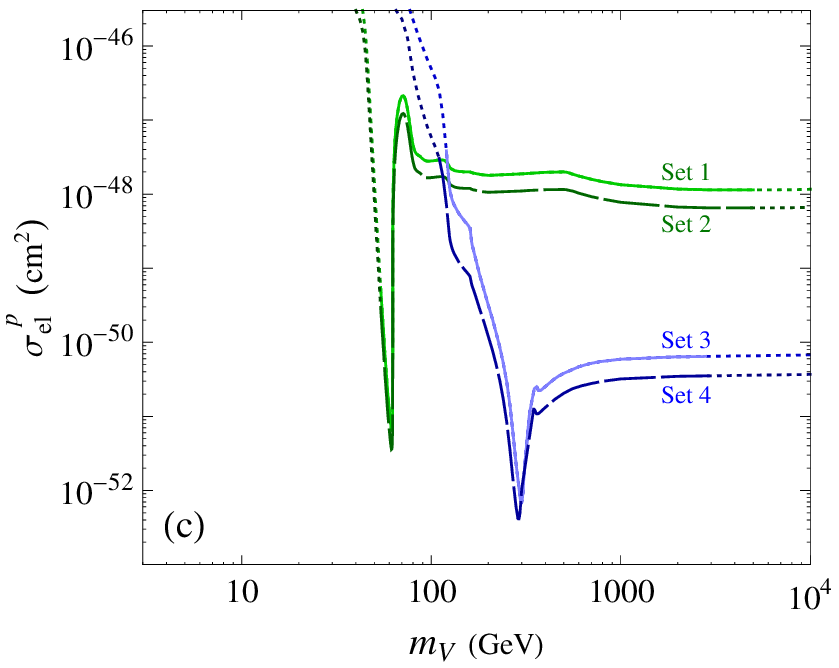}
\end{minipage} \vspace{-1ex}%
\caption{The same as Fig.\,\ref{2hdm+fdm_half_lambdas}, except the DM is the spin-1 singlet
\texttt{\slshape V} in the THDM II+V with (a)~$\lambda_H=0$ and (b)~$\lambda_h=0$.}
\label{2hdm+v_lambdas} \end{figure}
\begin{figure}[t]
\includegraphics[width=85mm]{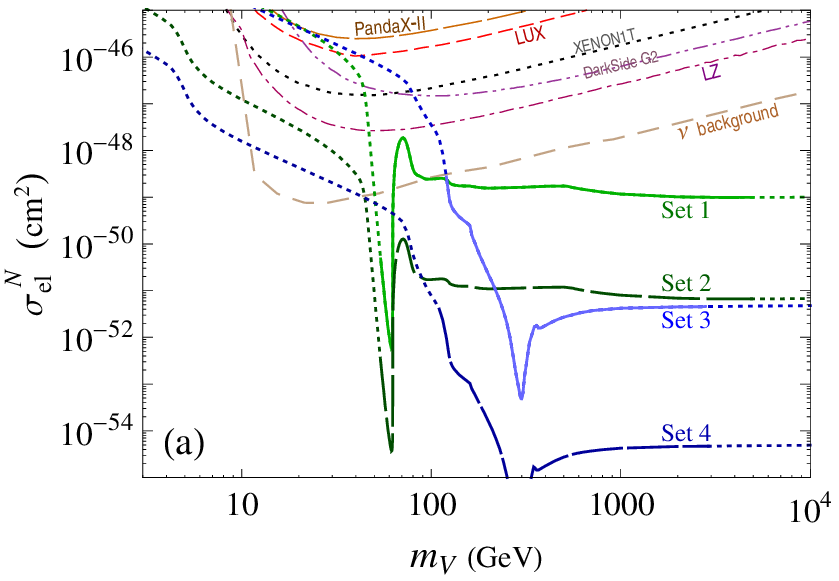}
\includegraphics[width=85mm]{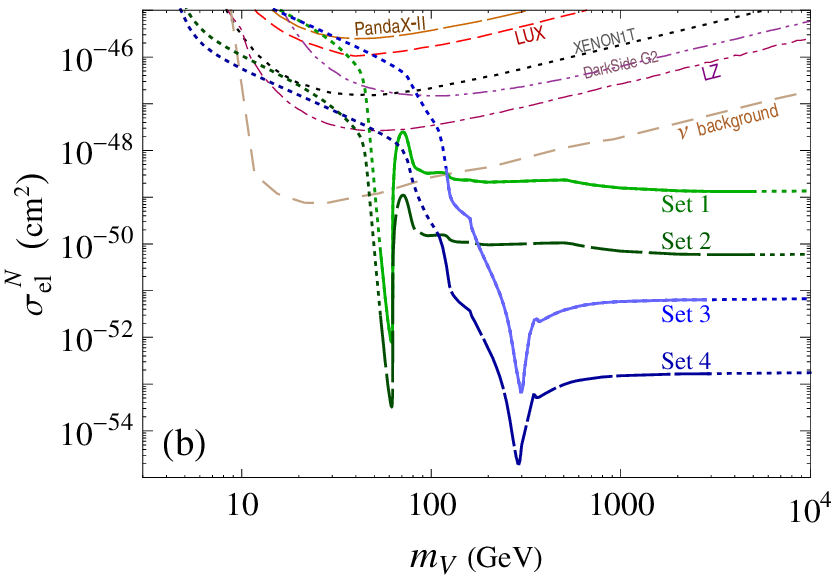}\vspace{-1ex}
\caption{The predicted DM-nucleon cross-sections (green curves) in the THDM II+V with input
numbers from Sets 1 and 2 (green curves) in Table\,\,\ref{lambdaH=0} and
Sets 3 and 4 (blue curves) in Tables\,\,\ref{lambdah=0} for (a)\,\,xenon and (b)\,\,argon targets,
compared to the same data and projections as in Fig.\,\,\ref{sm+fdm_half-plots}(b).
The dotted portions of the green and blue curves are disallowed as in
Fig.\,\,\ref{2hdm+v_lambdas}.}
\label{2hdm+v_sigmaelN} \end{figure}

Employing these formulas with the input numbers from Tables\,\,\ref{lambdaH=0}
and\,\,\ref{lambdah=0} for our examples, we arrive at the green and blue curves
in Figs.\,\,\ref{2hdm+v_lambdas} and\,\,\ref{2hdm+v_sigmaelN}.
In Fig.\,\,\ref{2hdm+v_lambdas}(a), we also draw the upper bound on $|\lambda_h|$ from
the Higgs data (black dotted curve).
As in the SM+V, theoretical considerations concerning unitarity and perturbativity lead to
the constraints \,$|\lambda_{\cal H}^{}|<\sqrt{2\pi}\,m_V^{}/v$\, and
\,$|\lambda_{\cal H}^{}|<1$,\, respectively, which are represented by the magenta dashed lines
in Fig.\,\,\ref{2hdm+v_lambdas}.
These instances illustrate that the situation in this model is roughly similar to that
in the THDM II plus real scalar DM investigated in Ref.\,\cite{He:2016mls}.
Specifically, in our $h$-portal examples the LHC constraint on
\,$h\to\texttt{\slshape V}\texttt{\slshape V}$\, rules out \,$m_V<54\,$GeV\,
and the effective theory is likely to be no longer perturbative for \,$m_V>5\,$TeV,\,
whereas in the $H$-portal cases the unitarity and perturbativity conditions disfavor
\,$m_V<110\,$GeV\, and \,$m_V>3\,$TeV.\,
Thus, more generally, in great contrast to the THDM II+$(\psi,\Psi)$ as well as the SM+V,
the THDM II+V has plentiful parameter space that is far away from the current direct
search limits and can even escape future ones.

\section{Conclusions\label{conclusion}}

The most recent limits from LUX and PandaX-II on DM-nucleon interactions and
the available LHC data on the 125-GeV Higgs boson's couplings have led to strong
restrictions on the simplest Higgs-portal dark matter models.
Taking these constraints into account, we have revisited the minimal models with
fermionic DM having spin 1/2 or 3/2 and a purely scalar
effective coupling to the Higgs doublet.
Realizing also that the EFT description for this coupling has limitations, we have
found that these minimal fermionic models are ruled out except in narrow regions of
the DM mass in the neighborhood of the Higgs resonance point at $m_h/2$.
On the other hand, the simplest Higgs-portal vector-DM model is viable not only
around \,$m_V=m_h/2$,\, but also above \,$m_V\sim1.4$\,TeV,\,
although it may lose perturbativity if \,$m_V>3.9\,$TeV.\,
Slightly expanding each of these models with the addition of another Higgs doublet and
assuming the extended Yukawa sector to be that of the type-II two-Higgs-doublet model,
we can significantly relax the restraints from direct search and LHC data,
even in the fermionic DM scenarios, and recover sizable parts of the regions
excluded in the minimal models.
This is due to suppression of the effective interactions of the portal $CP$-even Higgs bosons
with nucleons at some values of the ratios of the Higgs couplings to the up and down quarks,
rendering the interactions considerably isospin-violating.
Sizable portions of the revived parameter space can yield a DM-nucleon scattering
cross-section that is much smaller than its current experimental bound or even falls
under the neutrino-background floor.
Nevertheless, there are also areas in the parameter space of these two-Higgs-doublet-portal
DM models that are still within the discovery reach of future quests such as XENON1T and LZ.

\acknowledgments

This work was supported in part by MOE Academic Excellence Program (Grant No. 105R891505)
and NCTS of ROC.
X.-G. He was also supported in part by MOST of ROC (Grant No.~MOST104-2112-M-002-015-MY3)
and in part by NSFC (Grant Nos.~11175115 and 11575111), Key Laboratory for Particle Physics,
Astrophysics and Cosmology, Ministry of Education, and Shanghai Key Laboratory for Particle
Physics and Cosmology (SKLPPC) (Grant No.~11DZ2260700) of PRC.

\appendix

\section{Extra formulas for DM reactions\label{formulas}}

To extract the DM-Higgs coupling which enters the DM-annihilation cross-section
$\sigma_{\rm ann}$, we employ its thermal average~\cite{Gondolo:1990dk}
\begin{eqnarray} \label{sv}
\langle\sigma v_{\rm rel}^{}\rangle \,=\, \frac{x}{8 m_{\textsc{dm}}^{5\;}K_2^2(x)}
\int_{4m_{\textsc{dm}}^2}^\infty ds\;\sqrt s\,\bigl(s-4m_{\textsc{dm}}^2\bigr)\;
K_1^{}\bigl(\sqrt s\,x/m_{\textsc{dm}}^{}\bigr)\,\sigma_{\rm ann}^{} \,,
\end{eqnarray}
where $v_{\rm rel}$ is the relative speed of the DM pair, $K_r$ is the modified Bessel function
of the second kind of order $r$ and $x$ can be set to its freeze-out value \,$x=x_f^{}$,\,
which is related to $\langle\sigma v_{\rm rel}^{}\rangle$ by~\cite{Kolb:1990vq}
\begin{eqnarray} \label{xf}
x_f^{} \,=\, \ln\frac{0.038(2J_{\textsc{dm}}+1)m_{\textsc{dm}}^{}\,m_{\rm Pl}^{}\,
\langle\sigma v_{\rm rel}^{}\rangle}{\sqrt{g_*^{}x_f^{}}} \,,
\end{eqnarray}
with $J_{\textsc{dm}}$ being the DM particle's spin, \,$m_{\rm Pl}=1.22\times10^{19}$ GeV\,
the Planck mass, and $g_*^{}$ the number of effectively relativistic degrees of freedom below
the freeze-out temperature \,$T_f^{}=m_{\textsc{dm}}^{}/x_f^{}$.\,
In addition, we adopt the numerical values of $\langle\sigma v_{\rm rel}^{}\rangle$ versus DM
mass determined in Ref.~\cite{Steigman:2012nb},\footnote{Since our fermionic DM candidates
($\psi$ and $\Psi_\nu$) are complex fields, the $\langle\sigma v_{\rm rel}^{}\rangle$ values
used in (\ref{sv}) and (\ref{xf}) for the fermionic cases are twice those provided
by~\cite{Steigman:2012nb}, which applied to real or self-conjugate DM.}
as well as the latest relic density data \,$\Omega\hat h^2=0.1197\pm0.0022$\, \cite{planck},
with $\hat h$ being the Hubble parameter.

\begin{figure}[b]
\includegraphics[width=12cm]{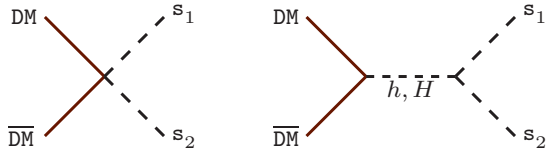}\vspace{-1ex}
\caption{Feynman diagrams contributing at leading order to DM annihilation into
${\texttt s}_1{\texttt s}_2$.} \label{dd2hh}
\end{figure}

In the THDM\,II+$\psi$, the DM annihilation mode \,$\bar\psi\psi\to{\texttt s}_1{\texttt s}_2$\,
can take place due to the diagrams displayed in Fig.\,\ref{dd2hh}, where
\,${\texttt s}_1{\texttt s}_2=hh,hH,HH,AA,H^+H^-$.\,
We ignore the contributions of $t$- and $u$-channel $\psi$-mediated diagrams because they
are at a higher order in $\lambda_{\psi\cal H}$ and of the same order as the potential
contributions of next-to-leading effective operators not included in Eq.\,(\ref{Lpsi'}).
The cross sections of these reactions are then calculated to be
\begin{eqnarray} \label{DD->hh}
\sigma\big(\bar\psi\psi\to{\texttt s}_1^{}{\texttt s}_2^{}\big) \,=\,
\frac{\beta_\psi^{~\;}\tilde\beta_{{\texttt s}_1{\texttt s}_2}^{}}
{32(1+\delta_{{\texttt s}_1{\texttt s}_2})\pi v^2}
\big( {\cal R}_{\psi,{\texttt s}_1{\texttt s}_2}^2
+ {\cal I}_{\psi,{\texttt s}_1{\texttt s}_2}^2 \big) \,,
\end{eqnarray}
where
\begin{eqnarray} \label{MXY}
\beta_{\textsc x}^{} &=& \sqrt{1-\frac{4m_{\textsc x}^2}{s}} \,, ~~~~ ~~
\tilde\beta_{{\texttt s}_1^{}{\texttt s}_2^{}}^{} \,=\,
\frac{\sqrt{\big(s-m_{{\texttt s}_1}^2-m_{{\texttt s}_2}^2\big)\raisebox{1pt}{$^2$}
- 4 m_{{\texttt s}_1\,}^2m_{{\texttt s}_2}^2}}{s} \,,
\nonumber \\
{\cal R}_{\psi,\textsc{xy}}^{} &=& \lambda_{\psi\textsc{xy}}^{} +
\frac{\lambda_{\psi h}^{}\lambda_{h\textsc{xy}}^{}\big(s-m_h^2\big)v^2}
{\big(s-m_h^2\big)\raisebox{1pt}{$^2$}+\Gamma_h^2m_h^2}
+ \frac{\lambda_{\psi H}^{}\lambda_{H\textsc{xy}}^{}\big(s-m_H^2\big)v^2}
{\big(s-m_H^2\big)\raisebox{1pt}{$^2$}+\Gamma_H^2m_H^2} \,,
\nonumber \\
{\cal I}_{\psi,\textsc{xy}}^{} &=&
\frac{\lambda_{\psi h}^{}\lambda_{h\textsc{xy}}^{}\Gamma_h^{}m_h^{}v^2}
{\big(s-m_h^2\big)\raisebox{1pt}{$^2$}+\Gamma_h^2m_h^2}
+ \frac{\lambda_{\psi H}^{}\lambda_{H\textsc{xy}}^{}\Gamma_H^{}m_H^{}v^2}
{\big(s-m_H^2\big)\raisebox{1pt}{$^2$}+\Gamma_H^2m_H^2} \,,
\end{eqnarray}
with $\lambda_{\psi\texttt{XY}}$ being connected to $\lambda_{\psi\cal H}$ for
\,${\cal H}=h,H$\, by Eq.\,\,(\ref{lambdahh}) and the expressions for the Higgs cubic
couplings $\lambda_{{\cal H}\textsc{xy}}$ listed in Ref.\,\cite{He:2016mls}.
In our scenarios of interest, $\Gamma_H$ gets contributions not only from the partial
rates of $H$ decays into fermions and gauge bosons, analogously to $\Gamma_h$,
but also from
\begin{eqnarray}
\Gamma\big(H\to\bar\psi\psi\big) \,=\, \frac{\lambda_{\psi H\,}^2m_H^{}}{8\pi}
\Bigg(1-\frac{4 m_\psi^2}{m_H^2}\Bigg)^{\!3/2} \,, ~~~ ~~~~
\Gamma(H\to hh) \,=\, \frac{\lambda_{hhH\,}^2v^2}{8\pi\,m_H^{}}
\sqrt{1-\frac{4 m_h^2}{m_H^2}}
\end{eqnarray}
once these channels are open.
The expression for $\sigma\big(\bar\psi\psi\to hh\big)$ in Eq.\,(\ref{DD->hh}) is applicable
to that in Eq.\,(\ref{DD2h2sm}) belonging to the SM+$\psi$, in which case there is only one
coupling for the $\psi$-Higgs interaction, \,$\lambda_{\psi hh}=\lambda_{\psi h}$,\,
and $H$ is absent, \,$\lambda_{\psi H}=\lambda_{hhH}=0$.\,

In the spin-3/2 DM model, THDM II+$\Psi$, the counterpart of Eq.\,(\ref{DD->hh}) is
\begin{eqnarray} \label{PP->hh}
\sigma(\bar{\Psi}\Psi\to{\texttt s}_1^{}{\texttt s}_2^{}) \,=\,
\frac{\big( 5\beta_\Psi^{}-6\beta_\Psi^3 +
9\beta_\Psi^5\big)\tilde\beta_{{\texttt s}_1{\texttt s}_2}^{}s^2}
{9216(1+\delta_{{\texttt s}_1{\texttt s}_2})\pi\,m_\Psi^4v^2}
\big( {\cal R}_{\Psi,{\texttt s}_1{\texttt s}_2}^2
+ {\cal I}_{\Psi,{\texttt s}_1{\texttt s}_2}^2 \big) \,,
\end{eqnarray}
where ${\cal R}_{\Psi,\textsc{xy}}$ $({\cal I}_{\Psi,\textsc{xy}})$ is the same as
${\cal R}_{\psi,\textsc{xy}}$ $({\cal I}_{\psi,\textsc{xy}})$ in Eq.\,(\ref{MXY}),
except the label $\psi$ in $\lambda_{\psi\cal H}$ and $\lambda_{\psi\textsc{xy}}$ is
replaced by $\Psi$.
The $\sigma(\bar{\Psi}\Psi\to hh)$ formula in Eq.\,(\ref{PP->hh}) becomes that in
Eq.\,(\ref{PP2h2sm}) belonging to the SM+$\Psi$ if we set
\,$\lambda_{\Psi hh}=\lambda_{\Psi h}$\, and \,$\lambda_{\Psi H}=\lambda_{hhH}=0$.\,

In the THDM II+V, the annihilation of the vector DM into a pair of Higgs bosons,
$\texttt{\slshape V}\texttt{\slshape V}\to{\texttt s}_1^{}{\texttt s}_2^{}$,\, is induced
by contact and $s$-channel diagrams (Fig.\,\ref{dd2hh}), as the $t$- and $u$-channel
ones are of higher order in $\lambda_{h,H}$ and hence neglected.
The cross section is then
\begin{eqnarray} \label{vv2ss'}
\sigma(\texttt{\slshape V}\texttt{\slshape V}\to{\texttt s}_1^{}{\texttt s}_2^{}) \,=\,
\frac{\big( \beta_V^2 s^2+12 m_V^4\big)\tilde\beta_{{\texttt s}_1{\texttt s}_2}^{}}
{144\beta_V^{}(1+\delta_{{\texttt s}_1{\texttt s}_2})\pi\,m_V^4 s}
\big({\cal R}_{{\texttt s}_1{\texttt s}_2}^2+{\cal I}_{{\texttt s}_1{\texttt s}_2}^2\big) \,,
\end{eqnarray}
where
\begin{eqnarray} \label{RI}
{\cal R}_{\textsc{xy}}^{} &=& \lambda_{\textsc{xy}}^{} +
\frac{\lambda_h^{}\lambda_{h\textsc{xy}}^{}\big(s-m_h^2\big)v^2}
{\big(s-m_h^2\big)\raisebox{1pt}{$^2$}+\Gamma_h^2m_h^2}
+ \frac{\lambda_H^{}\lambda_{H\textsc{xy}}^{}\big(s-m_H^2\big)v^2}
{\big(s-m_H^2\big)\raisebox{1pt}{$^2$}+\Gamma_H^2m_H^2} \,,
\nonumber \\
{\cal I}_{\textsc{xy}}^{} &=&
\frac{\lambda_h^{}\lambda_{h\textsc{xy}}^{}\Gamma_h^{}m_h^{}v^2}
{\big(s-m_h^2\big)\raisebox{1pt}{$^2$}+\Gamma_h^2m_h^2}
+ \frac{\lambda_H^{}\lambda_{H\textsc{xy}}^{}\Gamma_H^{}m_H^{}v^2}
{\big(s-m_H^2\big)\raisebox{1pt}{$^2$}+\Gamma_H^2m_H^2} \,.
\end{eqnarray}
The $\sigma(\texttt{\slshape V}\texttt{\slshape V}\to hh)$ expression in
Eq.\,(\ref{vv2ss'}) is applicable to the SM+V, in which case there is only one coupling
for the \texttt{\slshape V}-Higgs interaction, \,$\lambda_{hh}=\lambda_h$,\,
and $H$ is again absent, \,$\lambda_H=\lambda_{hhH}=0$.\,

\section{Conditions for perturbativity, vacuum stability, and tree-level
unitarity\label{theory-reqs}}

The parameters of the scalar potential \,${\cal V}_H$\, in Eq.\,\,(\ref{pot}) are subject
to a number of theoretical constraints.
The usual assumption that the scalar interactions are in the perturbative regime implies
that \,$|\lambda_{1,2,3,4,5}|\le8\pi$\, \cite{Kanemura:1999xf}.
The requisite stability of ${\cal V}_H$ implies that it has to be bounded from below,
entailing that
\begin{eqnarray} \label{stability}
\lambda_1^{} \,>\, 0 \,, ~~~ ~~~~
\lambda_2^{} \,>\, 0 \,, ~~~ ~~~~
\lambda_3 + {\rm min}(0,\lambda_4-|\lambda_5|) \,>\, -\sqrt{\lambda_1\lambda_2} \,.
\end{eqnarray}
Another important limitation is that the amplitudes for scalar-scalar scattering
\,$s_1^{}s_2^{}\to s_3^{}s_4^{}$\, at high energies respect unitarity.
This amounts to demanding that the combinations~\cite{Branco:2011iw,Kanemura:1993hm}
\begin{eqnarray} \label{unitarity} &
\tfrac{3}{2}(\lambda_1+\lambda_2) \pm
\sqrt{\tfrac{9}{4}(\lambda_1-\lambda_2)\raisebox{0.3pt}{$^2$}
+ (2\lambda_3+\lambda_4)\raisebox{0.3pt}{$^2$}} \,, ~~~~ ~~~
\tfrac{1}{2}(\lambda_1+\lambda_2) \pm
\sqrt{\tfrac{1}{4}(\lambda_1-\lambda_2)\raisebox{0.3pt}{$^2$}+\lambda_4^2} \,,
& \nonumber \\ &
\tfrac{1}{2}(\lambda_1+\lambda_2) \pm
\sqrt{\tfrac{1}{4}(\lambda_1-\lambda_2)\raisebox{0.3pt}{$^2$}+\lambda_5^2} \,, ~~~~ ~~~
\lambda_3^{} + 2 \lambda_4^{} \pm 3 \lambda_5^{} \,, ~~~~ ~~~
\lambda_3^{} \pm \lambda_4^{} \,, ~~~~ ~~~
\lambda_3^{} \pm \lambda_5^{} & ~~~~
\end{eqnarray}
each not exceed 8$\pi$ in magnitude.
We implement the conditions in Eqs.\,\,(\ref{stability}) and (\ref{unitarity})
employing the relations
\begin{eqnarray} \label{lambda1}
\lambda_1^{} &=& \frac{s_\alpha^2 m_h^2+c_\alpha^2 m_H^2}{c_\beta^{2\,}v^2}
- \frac{s_\beta^{} m_{12}^2}{c_\beta^{3\,}v^2} \,, ~~~ ~~
\lambda_3^{} \,=\, \frac{s_{2\alpha}^{}}{s_{2\beta}^{}}~\frac{m_H^2-m_h^2}{v^2}
+ \frac{2m_{H^\pm}^2}{v^2} - \frac{2 m_{12}^2}{s_{2\beta}^{}v^2} \,,
\nonumber \\
\lambda_2^{} &=& \frac{c_\alpha^2 m_h^2+s_\alpha^2 m_H^2}{s_\beta^{2\,}v^2}
- \frac{c_{\beta\,}^{}m_{12}^2}{s_\beta^{3\,}v^2} \,, ~~~~~
\lambda_4^{} \,=\, \frac{m_A^2-2m_{H^\pm}^2}{v^2} +
\frac{2 m_{12}^2}{s_{2\beta}^{}v^2} \,, ~~~~~
\lambda_5^{} = \frac{2 m_{12}^2}{s_{2\beta}^{}v^2} - \frac{m_A^2}{v^2}  ~~~~ ~~~
\end{eqnarray}
derived from ${\cal V}_H$.
Upon specifying $\alpha$ and $\beta$, we can then take $m_{h,H,A,H^\pm,12}$ and
$\lambda_{h,H}$ as the free parameters instead of $\lambda_{1,2,3,4,5}$.
The expressions in Eq.\,(\ref{lambda1}) are in agreement with those in
the literature~\cite{Gunion:2002zf}.

Since the vector-DM models in Secs.\,\,\ref{vdm} and\,\,\ref{2hdm+v} are not renormalizable
and violate unitarity, we need to impose unitarity restrictions on the DM couplings
$\lambda_{h,H}$.
Given that $\lambda_{h,H}$ and $\lambda_V^{}$ are free parameters in our analysis, for
simplicity we suppose that the DM self-coupling $\lambda_V^{}$ is absent~\cite{Lebedev:2011iq}.
The amplitude for
\,$\texttt{\slshape V}\texttt{\slshape V}\to\texttt{\slshape V}\texttt{\slshape V}$\,
at high energies, \,$\sqrt s\gg m_{V,h,H}$,\, is then
\begin{eqnarray}
{\cal M}_{\texttt{\slshape V}\texttt{\slshape V}\to\texttt{\slshape V}\texttt{\slshape V}}^{}
\,=\, \frac{4\big(\lambda_h^2+\lambda_H^2\big)v^2}{m_V^2} \,.
\end{eqnarray}
The unitarity condition implies that
\,$|{\cal M}_{\texttt{\slshape V}\texttt{\slshape V}\to\texttt{\slshape V}\texttt{\slshape V}}|
<8\pi$,\,
which translates into
\begin{eqnarray}
\lambda_h^2+\lambda_H^2 \;<\; \frac{2\pi\,m_V^2}{v^2} \,.
\end{eqnarray}
In the SM+V, this becomes \,$|\lambda_h|<\sqrt{2\pi}\,m_V^{}/v$,\, which is used
in Sec.\,\ref{vdm}.

There is a complementary theoretical restraint on $\lambda_h$ (or $\lambda_H$) having to do
with the supposed perturbativity of the effective \texttt{\slshape V} interactions.
To get a rough estimate on the implied cap on $\lambda_h$, we may assume that
\,$\lambda_h{\sf H}^\dagger{\sf H}\texttt{\slshape V}_\kappa\texttt{\slshape V}^{\,\kappa}$\,
in ${\cal L}_V$ arises from a tree-level diagram mediated by a heavy scalar $X$ coupled to
$h$ and \texttt{\slshape V} according to
\,${\cal L}_X^{}\supset-g_h^{}h^2X+g_V^{}
\texttt{\slshape V}_\kappa\texttt{\slshape V}^{\,\kappa}X$\, in the UV-complete theory,
as proposed in Ref.\,\cite{Lebedev:2011iq}.
We may further assume that \,$g_h^{}\sim\lambda_{hX}^{}v_X^{}$\, and
\,$g_V^{}\sim m_V^2/v_X^{}$,\, where $v_X^{}$ is the VEV of $X$, in analogy to the scalar and
weak-boson couplings in the SM, ignoring potential modifications due to $h$-$X$ mixing.
The EFT will then remain a reliable approach and perturbative if
\,$2|\lambda_h|\sim4|\lambda_{hX}|m_V^2/m_X^2<|\lambda_{hX}|<4\pi$,\,
as the $s$-channel energy $\sqrt s$ satisfies \,$m_X^2>s>4m_V^2$.\,
Since it is likely that the EFT description starts to break down at a lower $|\lambda_{hX}|$,
as was also suggested in Sec.\,\ref{sec:sm+fdm} for the fermionic cases, it is more reasonable to
select \,$|\lambda_h|<1$\, instead.
Similarly, we impose \,$|\lambda_H|<1$\, in the $H$-portal THDM\,\,II+V scenarios.

\end{document}